\documentclass[
 reprint,
 amsmath,amssymb,
 aps,superscriptaddress,
 nofootinbib
]{revtex4-2}

\usepackage{array}[=2016-10-06]

\usepackage{graphicx}
\usepackage{dcolumn}
\usepackage{bm}
\usepackage{mathtools}
\usepackage{siunitx}
\usepackage{hyperref}
\usepackage{braket}

\DeclarePairedDelimiterXPP\BigOSI[2]
  {\mathcal{O}}{(}{)}{}
  {\SI{#1}{#2}}

\hypersetup{
  colorlinks=true,
  linkcolor=blue,
  filecolor=magenta,
  citecolor=blue,
  urlcolor=blue
}

\newcommand\numberthis{\addtocounter{equation}{1}\tag{\theequation}}

\begin{document}

\preprint{APS/123-QED}

\title{Searching for binary black hole mergers with deep learning in Advanced LIGO's third observing run}

\author{Damon Beveridge}
 \email{damon.beveridge@uwa.edu.au}
\author{Alistair McLeod}
 \email{alistair.mcleod@research.uwa.edu.au}
\author{Linqing Wen}
 \email{linqing.wen@uwa.edu.au}
\author{Weichangfeng Guo}
\affiliation{Department of Physics, The University of Western Australia, 35 Stirling Highway, Crawley, WA 6009, Australia
 }

\author{Andreas Wicenec}
\affiliation{
International Centre for Radio Astronomy Research, The University of Western Australia
M468, 35 Stirling Hwy,  Crawley, WA 6009, Australia
}

\date{\today}

\begin{abstract}
The detection of gravitational waves from compact binary coalescences has provided significant insights into our Universe, and the discovery of new and unique gravitational wave candidates from independent searches remains an ongoing field of research. In this work, we built a hybrid search pipeline that combines matched filtering and deep learning to identify stellar-mass binary black hole candidates from detector strain data. We first present results from a targeted injection study to benchmark the sensitivity of our method and compare it with existing search pipelines. We demonstrate that our hybrid approach has comparable sensitivity for injections with a source-frame chirp mass greater than 25$\,$M$_{\odot}$, and below this threshold our sensitivity drops off for signals with a network SNR less than 15. We also observe that our search method can identify a significant population of unique candidates. Furthermore, we conduct an offline search for gravitational wave candidates in the third observing run of the LIGO-Virgo-KAGRA Collaboration (LVK), yielding 31 candidates previously reported by the LVK with a probability of astrophysical origin $p_{\rm astro}\geq0.5$. We identify two other candidates: one previously reported only in a search conducted by the Institute for Advanced Study, and one previously unreported promising new candidate with a $p_{\rm astro}$ of 0.63. This unique candidate has a high chirp mass and a high probability that the primary black hole is an intermediate-mass black hole.
\end{abstract}

\maketitle

\section{\label{sec:intro}Introduction}

The LIGO-Virgo-KAGRA Collaboration (LVK)~\cite{LVK} has ushered in an era of routine observations of gravitational waves (GWs) produced through compact binary mergers. Since the first detection in 2015, there have been published results from three completed observing runs (O1-O3) and one partial observing run (O4a) using the second-generation of gravitational-wave observatories, including the Hanford and Livingston LIGO observatories~\cite{LIGO}, Advanced Virgo~\cite{Virgo} and KAGRA~\cite{kagra}. Between observing runs, upgrades to the detectors have increased the rate of compact binary merger observations, resulting in over 218 observations at the end of O4a~\cite{gwtc1,gwtc2,gwtc21,gwtc3,gwtc40}. The majority of these candidates are the result of binary black hole (BBH) mergers, with two binary neutron star (BNS) mergers~\cite{gw170817, gw190425} and eight neutron star-black hole (NSBH) mergers~\cite{nsbh_detections,gw230529_181500,gwtc40,gwtc3,gwtc21,gw190814}. Since O4a, the fourth observing run (O4) has continued and has reported over 150 significant candidate events, as of the time of writing, in low-latency through the Gravitational-Wave Candidate Event Database (GraceDB)\footnote{\url{https://gracedb.ligo.org/}}, NASA's General Coordinates Network\footnote{\url{https://gcn.nasa.gov/}}, and the Scalable Cyberinfrastructure to support Multi-Messenger Astrophysics (SCiMMA) project\footnote{\url{https://scimma.org/}}~\cite{OPA}. These alerts encourage follow-up efforts for electromagnetic observation similar to how the first BNS event, GW170817, was observed by electromagnetic observatories~\cite{gw170817_em}. Increases in the observed populations of each compact binary source type and the observation of exceptional events, such as the massive GW190521 merger~\cite{gw190521} and the loud GW250114 merger~\cite{gw250114}, are starting to inform source type populations, merger rates and formation scenarios~\cite{2020ApJ...900L..13A,2021ApJ...913L..23E,2021NatAs...5..749G,2021ApJ...910..152Z}, and verify tests of general relativity~\cite{2021PhRvD.103l2002A}.

With the public release of LIGO and Virgo data, other groups have conducted offline analyses searching for GW signals~\cite{ogc1,ogc2,ogc3,ogc4,Nitz:single_det,Nitz:ecc_bns_o1o2,Nitz:high_q_ssm,Nitz:ssm_ecc,Nitz:ssm_o3a,Nitz:ssm_o3,pycbc_single_det,Dhurkunde:ecc_nsbh_bns_o3,Wang:bns_high_spin,Kacanja:ssm_bns_o3,pycbc_kde,ias_o1,ias_o2,ias_o3a,ias_o3b,ias_hom,ias_high_spin_o1,ias_disparate,cwb_ml,cwb_offline,gstlal_subthresh_bns,gstlal_gpu,gstlal_asym_precess_bbh_o3,Menendez-Vazquez:o2,Menendez-Vazquez:o3,aresgw,aframe,aframe_o3_search,gwak_offline}. These searches have reinforced the LVK's results and reported additional candidate events. The most notable of these catalogs of events are the Open Gravitational-Wave Catalog (OGC) series~\cite{ogc1,ogc2,ogc3,ogc4} and the Institute for Advanced Study (IAS) series~\cite{ias_o1,ias_o2,ias_o3a,ias_o3b,ias_hom}, which search over the first three observing runs for which public data is available. The additional searches generally introduce new ranking statistics for a given search pipeline that improve the search sensitivity, such as for the PyCBC-KDE~\cite{pycbc_kde} and cWB~\cite{cwb_offline} searches, or they introduce novel techniques for the method of detecting a merger; for example, the AresGW~\cite{aresgw} and Aframe~\cite{aframe_o3_search} searches use deep-learning models to identify BBH events. The AresGW (Aframe) search identified 52 (41) events in O3, including 39 (38) GWTC-3 candidates, 5 (3) candidates from other catalogs, and 8 (0) unique detections. These catalogs overlap significantly in the candidate events they identify, which are generally events in the LVK's Gravitational Wave Transient Catalogs (GWTC)~\cite{gwtc1,gwtc2,gwtc21,gwtc3}, and only 10 candidates have been identified by multiple independent searches outside of the GWTC candidate events in O3.

In this work, we build upon a prior feasibility study that combined matched filtering with a deep-learning model to identify BBH mergers from the signal-to-noise ratio (SNR) time series, using simulated Gaussian strain with injected glitches~\cite{Beveridge:bbh_gauss}. We previously demonstrated this method to be feasible for identifying BNS signals injected in O3 public data and the LVK candidate events GW170817 and GW190425~\cite{McLeod:bns}. Here, we apply the method proposed in Ref.~\cite{Beveridge:bbh_gauss} for the first time to detect BBH signals in O3 data from the LIGO Hanford and Livingston detectors, and evaluate the detection sensitivity of our search pipeline relative to existing LVK search pipelines. We present results from an offline search for candidate events in the publicly available O3 data, identifying 33 candidate events that exceed our detection threshold---defined as a probability of astrophysical origin, $p_{\text{astro}}$, greater than or equal to 0.5 and a false alarm rate (FAR) below 2 per day, which is consistent with the LVK Collaboration criteria~\cite{gwtc3}. There are 90 candidates in the GWTC-3 catalog; however, only 63 are considered detectable by the O3 offline search in this paper. Of the 33 detected candidates, 31 match those in the LVK's GWTC-3 catalog~\cite{gwtc3}, including the NSBH candidate GW190814. The remaining two candidates detected by our search include one that coincides with the IAS search that incorporates higher-order harmonics~\cite{ias_hom}, and one that represents a new candidate event not previously reported. We also demonstrate that our detected and missed candidates coincide with the populations of detected and missed injections from our sensitivity analyses, and we have no missed candidates that we expect to identify based on high SNRs and being within our most sensitive mass regions.

In Sec.~\ref{sec:method}, we discuss our search method and deep learning model in detail, including how it differs from Ref.~\cite{Beveridge:bbh_gauss}, and how we construct our FAR and $p_{\text{astro}}$ search statistics. In Sec.~\ref{sec:search-sens}, we apply our search pipeline to a set of signal injections placed into the O3 data and compare our sensitivity and recovered populations to the existing LVK search pipelines. Lastly, Sec.~\ref{sec:search} discusses the results of our offline search of O3 data, including the detected and missed populations of candidate events, and presents an analysis of our uniquely identified candidate and its inferred source parameters, as well as a discussion of areas for improvement in our search pipeline going forward.

\section{\label{sec:method}Method}

\begin{figure}[!t]
\includegraphics[width=\linewidth]{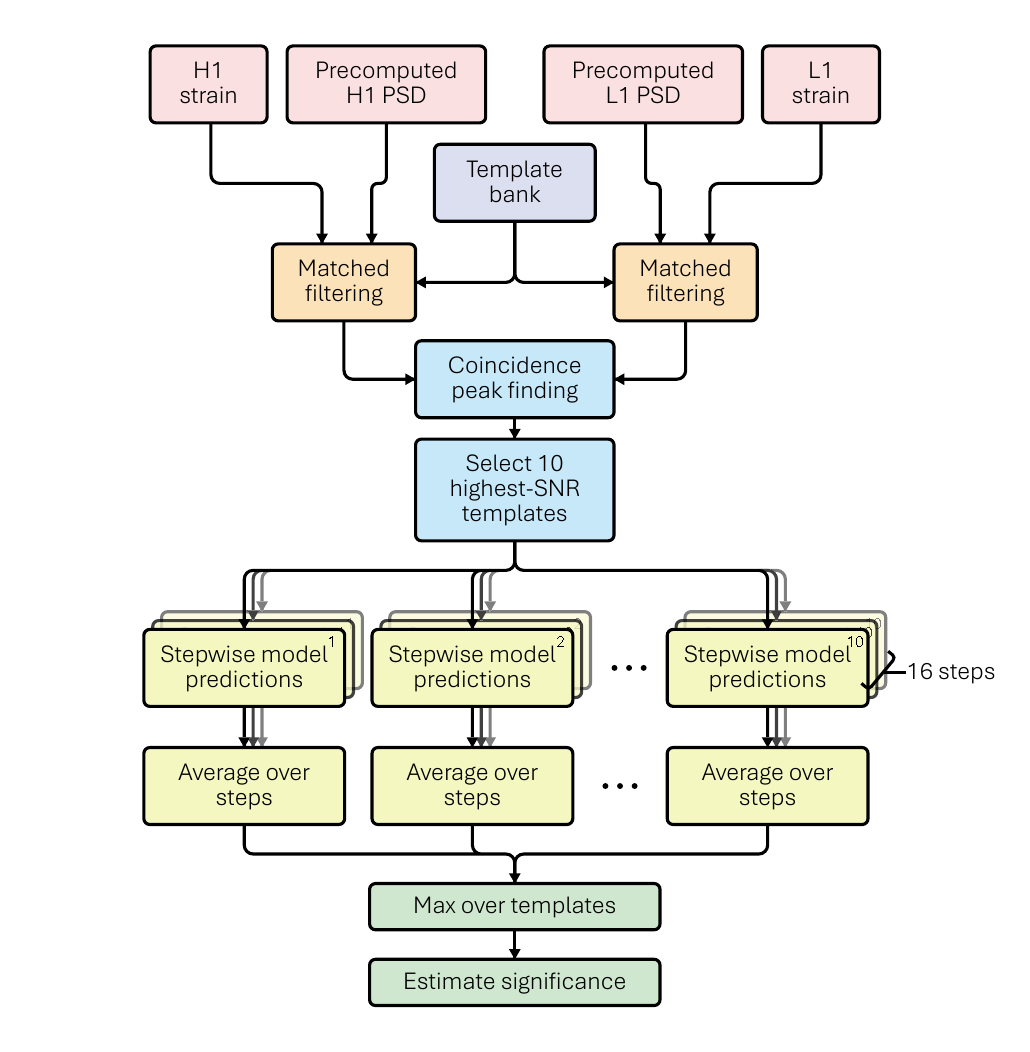}
\caption[Flowchart diagram for the search pipeline and ranking statistic]{\label{fig:flowchart}A flowchart of the search pipeline presented in this paper. The pipeline takes in public O3 data from the LIGO-Hanford (H1) and LIGO-Livingston (L1) detectors, as well as pre-computed PSDs of this data (Sec.~\ref{sec:data}). The data is downsampled to 2048$\,$Hz, and we perform matched filtering using a BBH template bank (Sec.~\ref{sec:template_bank}). A peak-finding algorithm then produces triggers based on SNR $>$ 4 peaks in at least one detector in the SNR time series, and identifies a coincident peak in the other detector within the light-travel time between detectors ($\sim$10$\,$ms). For each second of analyzed data, triggers are generated from the 10 highest network SNR templates, and the SNR time series is sent to our pre-trained deep learning model (Sec.~\ref{sec:training_dataset} and Sec.~\ref{sec:model_implementation}) for predictions. We perform 16 predictions by stepping through the SNR time series in increments of 1/16th of a second for each trigger. These stepwise predictions start when the trigger time first enters the deep learning model's 1-second viewing window and continue until it exits the window; the average of these predictions is then taken. We retain one of the 10 template triggers per second that has the highest average model prediction, and this quantity becomes the search pipeline's ranking statistic. The false alarm rate and probability of astrophysical origin can be estimated (Sec.~\ref{sec:search-stats}) by comparing the ranking statistic to empirical results from our search pipeline's analyses on time-shifted data and injections.}
\end{figure}

Our search method produces triggers using matched filtering, followed by a deep learning model that makes predictions on the SNR time series of these triggers. The ranking statistic is derived from these predictions and used to assign significance to a trigger. Matched filtering is the cross-correlation of detector strain data with a template waveform to produce a signal-to-noise ratio time series. In practice, an extensive collection of waveforms, known as a template bank, is used to search across the parameter space of compact binary objects. We have built a search pipeline that computes the SNR time series for our entire template bank, applies a template selection criterion using a peak-finding algorithm, and then uses a deep learning model to make predictions on the SNR time series and produce a ranking statistic and significance estimates. Figure~\ref{fig:flowchart} presents a flowchart of our search pipeline.

Section~\ref{sec:data} discusses the available data from the O3 observing run that we use for all training and analyses presented in this work. We discuss our chosen template bank in Section~\ref{sec:template_bank}. In Section~\ref{sec:training_dataset}, we discuss how we constructed the training dataset for our deep-learning model to achieve high sensitivity across the BBH parameter space. We discuss the implementation details of our deep-learning model further in Section~\ref{sec:model_implementation}. Lastly, we discuss how we compute our false alarm rate and probability of astrophysical origin metrics using our ranking statistic in Section~\ref{sec:search-stats}.

\subsection{\label{sec:data}Data}

We use public data from O3~\cite{open_data} for the Hanford and Livingston LIGO detectors, sampled at 2048$\,$Hz, for the model training and analyses presented herein. The O3 observing run consists of two distinct periods with a break in the middle for detector upgrades. The first period is O3a, running between 1 April 2019 15:00 UTC and 1 October 2019 15:00 UTC, and the second period is O3b, running from 1 November 2019 15:00 UTC to 27 March 2020 17:00 UTC. We require that both LIGO detectors have coincident data available and that each data segment have a minimum duration of 1024 seconds. This results in 106.2 days of data in O3a and 96.0 days in O3b. We do not use data from the Virgo detector due to its comparatively low sensitivity~\cite{gwtc3}.

For our offline search, we process the data in chunks of 1024 seconds. We perform matched filtering with a low frequency cutoff of 15$\,$Hz and do not produce triggers in the first 200$\,$s of each analysis segment to avoid filter wraparound effects for the longest waveforms in the template bank~\cite{findchirp}. We also do not produce triggers in the final 24$\,$s of each analysis segment to account for edge effects from whitening with the power spectral density (PSD). These cuts reduce our livetime for sensitivity tests and offline search to 104.6 days in O3a and 94.7 days in O3b.

To maintain consistency in our analyses and account for changes in PSDs over time, we aggregate data by calendar weeks of the O3 timeline. We compute the PSDs over the entire week of analyzed data, and the PSD used for matched filtering is updated every week to be the PSD of that week's data. No data is available between weeks 27 and 31 due to the break between O3a and O3b. Additionally, weeks 27, 31, and 52 have reduced observing time because the observing runs do not start or end on calendar-week boundaries.

\subsection{\label{sec:template_bank}Template bank}

The template bank used for this work is a subset of the templates from GstLAL's second observing run template bank~\cite{gstlal_o2_bank}, serving as a simple starting point for our search pipeline. The SPIIR pipeline also used this template bank for the real-time search during O3 \cite{QiChu2022}. We use a subset of the template bank in which both component masses exceed 2$\,$M$_\odot$, yielding 97,802 templates. This parameter space for our BBH search is selected based on the population of injections defined in the GWTC-3 BBH set of injections~\cite{gwtc3,gwtc3_inj_set}. We use the SEOBNRv4\_ROM waveform approximant~\cite{Bohe2017} with $f_{\rm low}$ = 15$\,$Hz for our matched-filter template waveforms.

\subsection{\label{sec:training_dataset}Training dataset}

The training dataset balances SNR time series samples of pure noise, glitch-contaminated noise, and injected BBH signals to expose the deep-learning model to realistic noise backgrounds and signals. We use strain data from week 1 of the O3 observing run for the noise background. We pair data from each detector by offsetting the strain in time by an interval longer than the light travel time between the detectors ($\sim$10$\,$ms), ensuring the model never sees coincident detector data during training. The strain data and subsequent SNR time series have a sampling rate of 2048$\,$Hz, and the training sample length is one second.

The background noise exhibits non-stationary, non-Gaussian behaviour. Of this behavior, the primary concern is the occurrence of frequent, short-duration glitches that can lead to false triggers in search pipelines~\cite{Christensen_2004,gracedb}. For this work, we use the existing Omicron software package~\cite{omicron} to identify transient noise events, enabling us to select data segments with and without glitches for use in our training dataset. During training, we only consider glitches identified by Omicron with an SNR greater than 6. Following the method in Ref.~\cite{McLeod:bns}, injected signals are placed relative to glitches such that the frequency content of the glitch aligns with the corresponding frequency in the injected waveform, ensuring that the glitch and the signal contribute to the matched filter SNR at the same time, maximizing their interaction. To prevent the glitch from always appearing centered in the SNR time series, we apply a random temporal shift of up to $\pm0.5$ seconds to the injected signal's position. Compared to our past work, we increase the proportion of samples containing glitches to $30\%$ to both improve model behavior in the presence of glitches and account for the random timing of glitches.

The training dataset contains 100,000 strain samples with only noise and 100,000 strain samples with injected BBH signals added to the noise. During training, we reserve 20\% of these samples for validation. As discussed previously, we pair each strain sample with 10 templates, with template selection governed by distinct strategies for noise-only and injection samples. Noise-only samples use randomly chosen templates, with each template having an equal chance. For injection samples, a set of 500 templates is initially selected by sorting the template bank by chirp mass, and taking 250 templates from either side of the injection's chirp mass. The match — the maximized overlap — is then calculated for each template using the injection waveform. Finally, the highest-matching template and nine others, randomly selected with a match greater than 0.5, are chosen to produce the training samples with injected signals. This allows the model to be trained on high- and low-match scenarios, which can occur in real detections.

\begin{figure}[t]
\includegraphics[width=\linewidth]{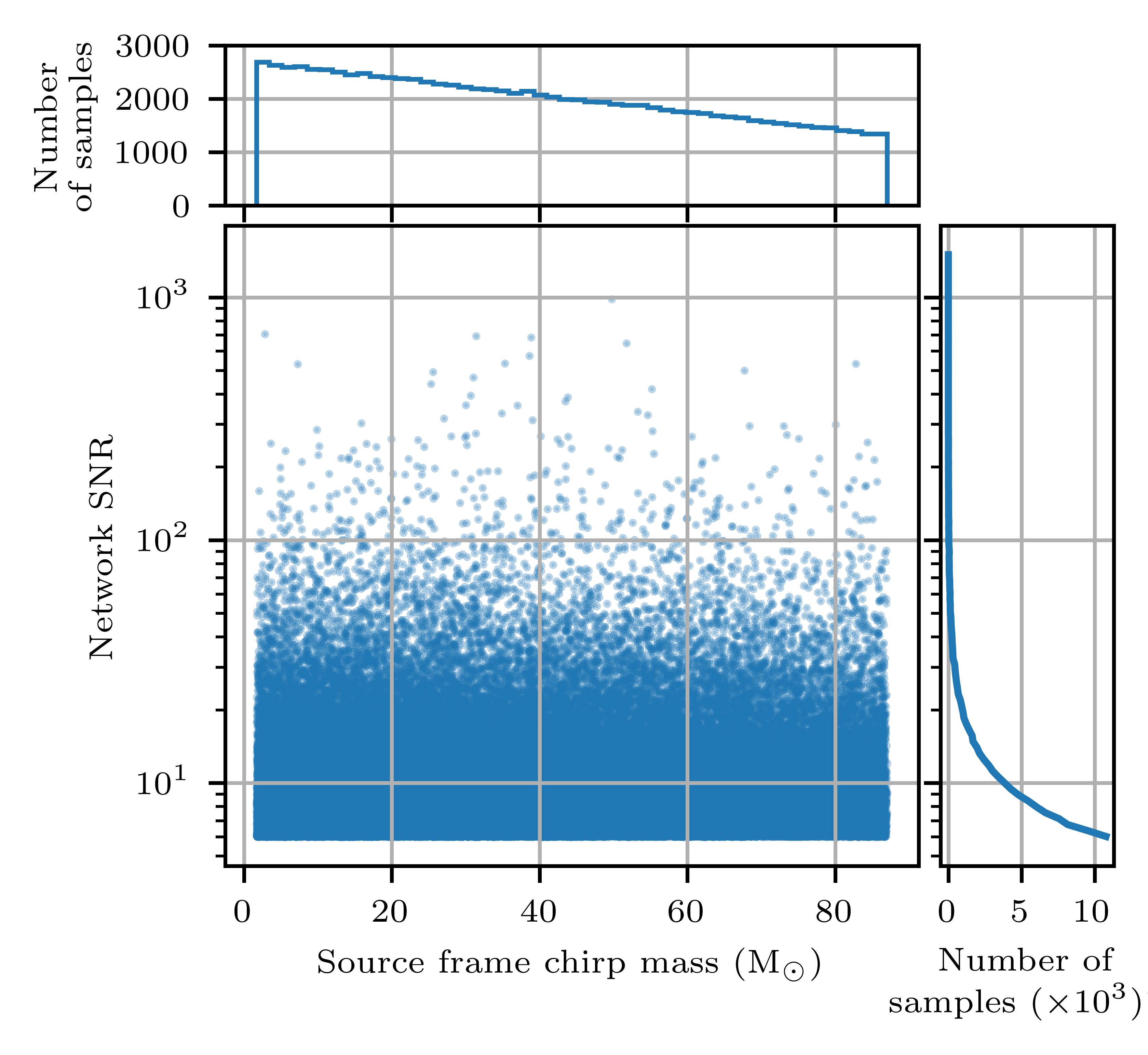}
\caption[Distribution of training dataset injections in source frame chirp mass and network SNR]{\label{fig:train_dataset}Source frame chirp mass and network signal-to-noise ratio (SNR) parameter distributions for the training dataset injection samples. The top histogram shows a bias towards low-mass signals, with twice as many samples in the lowest mass bin as in the highest. The histogram on the right shows the power-law distribution of network SNR. We chose these distributions to account for low injection counts in the high-mass and high-SNR regions when sampling from an astrophysical distribution.}
\end{figure}

For the set of injected signals, we optimize the parameter distributions to most effectively train our model to detect signals across the parameter space we search. In Ref.~\cite{Beveridge:bbh_gauss}, we found that using astrophysical priors for all injected signal parameters in the training dataset resulted in low sensitivity at higher masses, due to the astrophysical distribution being sharply biased toward low masses and containing very few samples at higher masses. We counteract this by manually controlling the density of injections in the source frame chirp mass. We choose source frame component masses to be between 2 and 100$\,$M$_\odot$, meaning the chirp mass has a lower bound corresponding to a 2-2$\,$M$_\odot$ pair and an upper bound of a 100-100$\,$M$_\odot$ pair. This range is chosen based on the selected astrophysical parameter space for binary black hole mergers in the LVK's GWTC-3 catalog~\cite{gwtc3,gwtc3_inj_set}, despite the low-mass region being more likely to represent mergers containing one or more neutron stars. To control the distribution of injections in chirp mass, we sample $\mathcal{O}(10^7)$ injections from astrophysical priors matching those in the GWTC-3 injection set~\cite{gwtc3_inj_set}, and sample from this set to construct a training subset with specific target distributions. The target distribution is linear and decreasing in chirp mass, with twice as many events in the lowest chirp mass region as in the highest, as shown in Figure~\ref{fig:train_dataset}. To achieve this distribution, we divide the source frame chirp mass parameter range into 50 uniformly sized bins. We then compute the required number of samples per bin and randomly sample the required number of injections from the set of $\mathcal{O}(10^7)$ injections for each bin. Following this sampling, we adjust the luminosity distances of each sample to achieve a target expected network SNR, ensuring the distribution follows a power law with an exponent of $-3$, with lower and upper bounds of 6 and 1000. Here, expected SNR refers to the optimal matched-filter SNRs for a simulated signal. The resulting dataset preserves astrophysical distributions for the intrinsic and extrinsic parameters unrelated to masses and distances.

Injection waveforms are generated using the SEOBNRv4PHM model~\cite{seobnrv4phm}, or the SEOBNRv4P model~\cite{seobnrv4phm} if the binary has a detector frame total mass below 9$\,$M$_\odot$, as in GWTC-3~\cite{gwtc3}. Higher-order modes are excluded at these lower masses because they would lie above the Nyquist frequency of the strain data. We position injection merger times so they fall within the central 0.9 seconds of the resulting 1-second SNR training sample, thereby training the model to identify signals across its input window.

\subsection{\label{sec:model_implementation}Model implementation}

The model is constructed and trained using \textsc{TensorFlow}~\cite{tensorflow2015}, and the model architecture is the same as was used in Ref.~\cite{Beveridge:bbh_gauss}. The model takes as input the SNR time series from both detectors and consists of a series of convolutional \cite{cnn}, residual (ResNet) \cite{he2015deep}, and fully connected layers that process each detector's data separately before combining them to form a single model output. During training, the model employs a sigmoid activation function on the output to constrain it to $0-1$, enabling the use of the binary cross-entropy loss function and accuracy as training metrics. Our model employs 32-bit floating-point precision and is susceptible to numerical overflow and underflow (rounding to 1 or 0). As a result, we remove the sigmoid activation function after training, resulting in an unbounded output that we use to compute the ranking statistic for our search.

After training the model and removing the sigmoid function on the model output, we accelerate its inference by converting it to an ONNX model with \textsc{ONNX Runtime}~\cite{onnxruntime}. Using ONNX allows us to run our offline search without a GPU, as CPU-based predictions are efficient enough to keep up with the other search pipeline components, and CPUs were more accessible than GPUs at the time of the analyses presented in this work.

\subsection{\label{sec:search-stats}Search statistics}

As mentioned at the start of Section~\ref{sec:method}, our search pipeline's ranking statistic is the highest average prediction from our deep-learning model on the SNR time series from the ten templates with the highest SNR over a given second, limiting our trigger rate to one per second. The purpose of the ranking statistic is to assign significance to a trigger, thereby classifying it as arising from noise or a gravitational wave signal based on a specific threshold. Gravitational wave detection uses two main statistics for classification: the false alarm rate (FAR) and the probability of astrophysical origin ($p_{\rm astro}$).

\subsubsection{\label{sec:fars}False alarm rate}

The false alarm rate (FAR) is an approximate measure of how often a search would produce a trigger, assuming it were caused by noise, with a ranking statistic, $\mathcal{R}$, greater than that of the trigger. We compute the FAR by collecting a background distribution of the ranking statistic and examining the rate at which these noise triggers occur; this allows us to assign an FAR to future triggers produced by the search. Currently, the false alarm rate is used as the public alert threshold for significant compact binary coalescence events during LVK observing runs, with any triggers with an FAR below 1 per month released to the public during O4~\cite{Chaudhary:2023vec}.

To scale up the amount of available background data, we decouple the strain from each detector by shifting them in time by an amount greater than the light travel time between the detectors. This shift results in a noise dataset with no coincident astrophysical events, but non-stationary noise artifacts remain, so that it can be treated as pure noise by most searches. Additionally, applying these shifts multiple times to the same strain segments multiplies the available noise data, meaning relatively little data is needed to surpass the FAR thresholds commonly used for public alerts. When collecting background, we shift the noise background data 100 times by increments greater than the light travel time, to achieve $\sim$1.5 years of collected background data from one week of O3 data.

Since our search produces triggers when at least one detector has an SNR greater than 4, we remove times of strain from the noise background that include candidate events in the GWTC-3 catalog to avoid contaminating the background. When analyzing our pipeline's performance, we identified drift in the distribution of the ranking statistic over time, so we regularly update our background collection as follows. At the start of O3a, we collect background on weeks 2 and 3, since we trained on week 1, and each has less Hanford-Livingston coincident data than desired. For the start of O3b, we use weeks 31 and 32, where week 31 is the beginning of O3b, but has less strain than desired to collect enough background. Our background data is then updated at regular intervals on weeks 5, 9, 13, 17, 21, and 25 in O3a, and on weeks 35, 39, 43, 47, and 51 in O3b. We use the updated background data to assign FARs in the same week it is collected, and every week until the following week that we collect new background data.

Due to the limited quantity of collected background each time we update it, we fit a linear function to the tail of the background data and use this to assign continuous FAR values below 1 per 2 months. We compute the fitted line as the weighted average of 10 fits from 10 random, independent subsets of the collected background. The coefficient of determination, ${\rm R}^2$, of each fit weights the slope and offset components to compute the overall fit.

\subsubsection{\label{sec:pastro}Probability of astrophysical origin}

While the false alarm rate gives us a measure of the expected rate of a trigger if it were to come from noise, it does not provide a direct measure of how likely a candidate trigger is to originate from an astrophysical source. The probability of astrophysical origin, $p_{\rm astro}$, estimates the probability that a candidate trigger represents a real gravitational wave signal. In contrast to the false alarm rate, the probability of astrophysical origin incorporates knowledge of the astrophysical signal rate and population.

The original method for computing $p_{\rm astro}$ is the FGMC method~\cite{Farr:2013yna}. It constructs a two-component model that assumes only two event types exist: noise (background) and signal (foreground). If we consider only triggers above a threshold where the background dominates the population of triggers, we can model the foreground and background trigger rates as two independent Poisson processes. For this, we consider triggers below an FAR threshold of $1\times10^{-4}\,$Hz. Since we treat our pipeline as only searching for binary black hole mergers, we do not introduce additional complexity into the model by computing the $p_{\rm astro}$ for multiple source types. We also tune separate $p_{\rm astro}$ models for the O3a and O3b observing runs; however, the tuning method is identical.

Our $p_{\rm astro}$ implementation is based on the implementation provided by the LVK's \textsc{p-astro}\footnote{https://git.ligo.org/lscsoft/p-astro} package, and is the same method the SPIIR pipeline~\cite{QiChu2022} uses to produce $p_{\rm astro}$ estimates for real-time detections as of the time of writing. To compute the $p_{\rm astro}$, we use a signal-versus-noise Bayes factor, defined as the ratio of analytically estimated foreground and background likelihood distributions. Given a set of N candidate events $\overrightarrow{x}_N = \{x_0, x_1,\ldots,x_{N-1}\}$, we can compute the mean values for the background Poisson count, $\braket{\Lambda_0}_N$, and the foreground Poisson count, $\braket{\Lambda_1}_N$. The probability of astrophysical origin of a new candidate, $x_{N+1}$, is~\cite{Kapadia:2019uut}:
\begin{equation}\label{eq:pastro}
P_1(x_{N+1}|\overrightarrow{x}_{N+1}) = \frac{\braket{\Lambda_1}_N K(x_{N+1})}{\braket{\Lambda_0}_N + \braket{\Lambda_1}_N K(x_{N+1})},
\end{equation}
where $x$ are the ranking statistics associated with a candidate, and $K(x)$ is the signal-versus-noise Bayes factor.

Following the approach in \textsc{GWCelery} \cite{gwcelery}, we model the background probability density, $p_0(\mathcal{R})$, in terms of the ranking statistic $\mathcal{R}$,
\begin{equation}\label{eq:pastro_bg}
p_0(\mathcal{R}) = \frac{\mathcal{F}(\mathcal{R})}{\mathcal{F}_{\rm th}},
\end{equation}
where $\mathcal{F}(\mathcal{R})$ is the false alarm rate expressed as a function of the ranking statistic, and $\mathcal{F}_{\rm th}$ is a threshold FAR that is tuned for the search. Here, $\mathcal{F}(\mathcal{R})$ is a cumulative quantity that is proportional to a probability density under the assumption that the ranking-statistic distribution is well-approximated by an exponential at low false alarm rates, $\mathcal{F}(\mathcal{R}) \propto {\rm exp}(-\alpha \mathcal{R})$ \cite{McLeod:bns,Usman:2015kfa}. The overall normalization is set by $\mathcal{F}_{\rm th}$, and any residual constant from the exponential approximation is absorbed into this normalization.

The foreground distribution, $p_1(\rho)$, represents the likelihood that a signal is observed with an SNR $\rho$, given that a signal is detected above an SNR threshold $\rho_{\rm th}$. Assuming a distribution of sources that is uniform in volume, the foreground distribution is~\cite{Schutz:2011tw, Chen:2014yla, LIGOScientific:2016ebi}:
\begin{equation}\label{eq:pastro_fg}
p_1(\rho) = \frac{3\rho_{\rm th}^3}{\rho^4},
\end{equation}
where $\rho_{\rm th}$ is a threshold SNR value that has to be tuned. From these analytic estimates, the Bayes factor, $K(\rho,\mathcal{R})$ can be defined as:
\begin{align*}
K(\rho,\mathcal{R}) &= \frac{p_1(\rho)}{p_0(\mathcal{R})} \\
                    &= \frac{3\rho_{\rm th}^3\mathcal{F}_{\rm th}}{\rho^4\mathcal{F}(\mathcal{R})}. \numberthis \label{eq:pastro_bayes_factor}
\end{align*}

\begin{figure}[t]
\includegraphics[width=\linewidth]{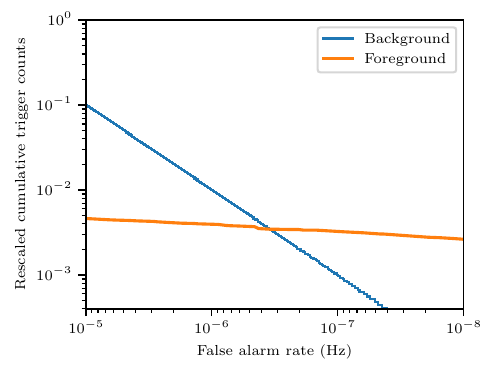}
\caption[Cumulative trigger counts for background and foreground events used to tune the $p_{\rm astro}$ model]{\label{fig:pastro_trig_densities}Cumulative trigger counts from O3a for the foreground (orange) and background (blue) distributions as a function of false alarm rate, scaled by their FGMC-inferred Poisson counts $\braket{\Lambda_0}_N$ and $\braket{\Lambda_1}_N$. We tune the $\mathcal{F}_{\rm th}$ and $\rho_{\rm th}$ in our Bayes factor such that the FAR at the intersection of the two distributions corresponds to a $p_{\rm astro}$ of 0.5.}
\end{figure}

\begin{figure}[t]
\includegraphics[width=\linewidth]{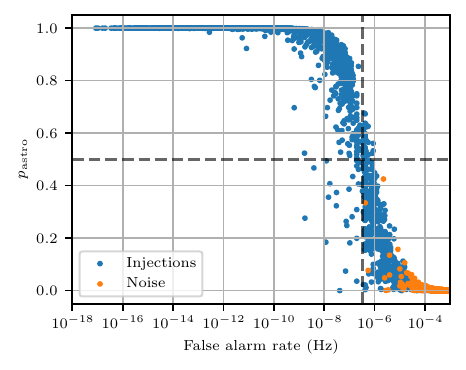}
\caption[Distribution of $p_{\rm astro}$ versus false alarm rate values for injection and noise triggers from an injection run]{\label{fig:pastro_vals}Probability of astrophysical origin ($p_{\rm astro}$) results on injection runs in weeks 18 and 20 of O3. We observe that a $p_{\rm astro}$ of 0.5 (horizontal black dotted line) correlates roughly to the cross-over FAR from Figure~\ref{fig:pastro_trig_densities} (vertical black dotted line).}
\end{figure}

In principle, a probability of astrophysical origin of 0.5 should indicate a trigger at a ranking statistic value where the trigger densities of background and foreground triggers are equal. This feature is how we tune the values of $\rho_{\rm th}$ and $\mathcal{F}_{\rm th}$, using empirical trigger densities of the background and foreground. Here, the background empirical distribution combines all background points from time-shifted triggers for a given observing run, as outlined in Section~\ref{sec:fars}. We estimate the foreground empirical distribution from the triggers produced in the first three weeks of injection run testing in each half of the O3 observing run, and the specific weeks are outlined in Section~\ref{sec:search-sens}. Figure~\ref{fig:pastro_trig_densities} presents the foreground and background cumulative trigger counts as a function of FAR, scaled by the Poisson counts $\braket{\Lambda_0}_N$ and $\braket{\Lambda_1}_N$ respectively, for the O3a observing run. The Poisson counts are inferred via the FGMC mixture formalism \cite{Farr:2013yna}. We tune the $\rho_{\rm th}$ and $\mathcal{F}_{\rm th}$ in the $p_{\rm astro}$ model such that for a test dataset of injection run results, the FAR value where the rescaled cumulative trigger counts intersect approximately lines up with a $p_{\rm astro}$ of 0.5, as can be seen in Figure~\ref{fig:pastro_vals}. Figure~\ref{fig:pastro_vals} represents the case in O3a, where the injections used are from weeks 18 and 20 using the GWTC-3 binary black hole injection set~\cite{gwtc3_inj_set}. Significant deviations from the S-curve in Fig.~\ref{fig:pastro_vals} mainly occur when injections have low likelihood in the deep learning model training data set or the astrophysical BBH population used to tune the $p_{\rm astro}$ model. These cases - typically injections with high component masses or masses consistent with an NSBH source - result in a less well-constrained relationship between FAR and SNR, yielding uncertain $p_{\rm astro}$ values. For tuning the O3b $p_{\rm astro}$, we test the model on injections from weeks 48 and 50. The tuned values for $\{\rho_{\rm th}, \mathcal{F}_{\rm th}\}$ are \{8.5, $10^{-4}$\} for O3a, and \{7.25, $10^{-4}$\} for O3b.

\section{\label{sec:search-sens}Search sensitivity}

To compare our search pipeline with existing pipelines, we quantify search sensitivity in terms of the spacetime volume in which signals can be detected - the sensitive hypervolume $\braket{VT}$ - using injected gravitational wave signals throughout the O3 observing run. The set of injected signals we use is from the BBH-only injection sets provided by the LVK with the third gravitational wave transient catalog~\cite{gwtc3_inj_set}. Using this dataset enables us to compare the sensitivity of our search with that of the existing cWB, GstLAL, MBTA, PyCBC-broad and PyCBC-BBH searches, as the $p_{\rm astro}$ and FAR results from these searches are provided. We treat the PyCBC-broad and PyCBC-BBH pipelines as independent; however, they differ only in the template bank parameter space (BNS, NSBH, and BBH versus just BBH), and the PyCBC-BBH pipeline ranking statistic incorporates a chirp mass weighting for triggers~\cite{gwtc3,ogc2}. The sensitivity comparisons we present serve only as a general benchmark, as each search pipeline is developed with a different search space and tuned towards sensitivities in various areas of the shared parameter space.

For this work, we search for injected signals every other week, starting in week 4 in O3a and in week 32 in O3b. The result is 83,815 binary black hole signal injections. We have compared sensitivity results for this set of injections with the following search pipelines \cite{gwtc3,gwtc3_inj_set}: PyCBC-broad, PyCBC-BBH, GstLAL, MBTA, and cWB. Note that these other searches have searched for injections in other weeks and with different online detector configurations \cite{gwtc3,gwtc3_inj_set}, which would increase the reported $\braket{VT}$ due to the increased livetime of the analyzed data. Additionally, each search pipeline has its own template bank, each with varying parameter spaces. However, for this work, we consider only the sensitivity of the other searches to the same BBH injections we are searching for, aiming for a more direct comparison of sensitivity before developing single-detector search capabilities and two- and three-detector searches involving Virgo. Extending our search to detect events across the full BNS and NSBH parameter spaces is expected to reduce sensitivity in the BBH region, thereby changing the population of noise events produced by our search.

{\renewcommand{\arraystretch}{1.5}%
\begin{table*}[]
\begin{ruledtabular}
\begin{tabular}{c cccccc cc}
\multicolumn{1}{c}{ } & \multicolumn{8}{c}{Sensitive hypervolume (Gpc$^3$ yr)}\\
\cline{2-9}
$m_1$, $m_2$ & cWB & GstLAL & MBTA & PyCBC-broad & PyCBC-BBH & This Work & Combined$^*$ & Combined$^\dagger$ \\
\hline
50, 50 & 3.51$^{+0.08}_{-0.07}$                & 4.14$^{+0.09}_{-0.08}$                & 3.78$^{+0.08}_{-0.08}$                & 3.35$^{+0.07}_{-0.08}$                & 4.49$^{+0.09}_{-0.08}$                & 4.00$^{+0.08}_{-0.09}$                & 5.33$^{+0.09}_{-0.10}$                & 5.81$^{+0.09}_{-0.10}$\\
50, 35 & 2.71$^{+0.07}_{-0.06}$                & 3.22$^{+0.07}_{-0.07}$                & 2.98$^{+0.06}_{-0.07}$                & 2.69$^{+0.07}_{-0.06}$                & 3.68$^{+0.07}_{-0.08}$                & 3.17$^{+0.07}_{-0.07}$                & 4.25$^{+0.08}_{-0.08}$                & 4.65$^{+0.08}_{-0.09}$\\
50, 20 & 1.18$^{+0.07}_{-0.08}$                & 1.61$^{+0.09}_{-0.09}$                & 1.43$^{+0.08}_{-0.09}$                & 1.37$^{+0.08}_{-0.09}$                & 1.89$^{+0.10}_{-0.09}$                & 1.35$^{+0.08}_{-0.08}$                & 2.14$^{+0.11}_{-0.10}$                & 2.34$^{+0.10}_{-0.11}$\\
35, 35 & 1.76$^{+0.04}_{-0.04}$                & 2.39$^{+0.05}_{-0.05}$                & 2.22$^{+0.05}_{-0.04}$                & 2.09$^{+0.05}_{-0.05}$                & 2.77$^{+0.06}_{-0.05}$                & 2.25$^{+0.05}_{-0.05}$                & 3.13$^{+0.06}_{-0.06}$                & 3.43$^{+0.06}_{-0.07}$\\
35, 20 & 0.90$^{+0.04}_{-0.04}$                & 1.41$^{+0.05}_{-0.05}$                & 1.21$^{+0.05}_{-0.04}$                & 1.24$^{+0.05}_{-0.04}$                & 1.65$^{+0.05}_{-0.05}$                & 1.16$^{+0.04}_{-0.04}$                & 1.82$^{+0.05}_{-0.06}$                & 1.95$^{+0.06}_{-0.05}$\\
20, 20 & 0.39$^{+0.01}_{-0.02}$                & 0.76$^{+0.02}_{-0.02}$                & 0.75$^{+0.02}_{-0.02}$                & 0.73$^{+0.02}_{-0.02}$                & 0.91$^{+0.02}_{-0.03}$                & 0.52$^{+0.02}_{-0.02}$                & 0.99$^{+0.03}_{-0.02}$                & 1.03$^{+0.02}_{-0.02}$\\
20, 10 & 0.17$^{+0.01}_{-0.02}$                & 0.36$^{+0.02}_{-0.02}$                & 0.35$^{+0.01}_{-0.02}$                & 0.36$^{+0.02}_{-0.02}$                & 0.41$^{+0.02}_{-0.02}$                & 0.21$^{+0.02}_{-0.01}$                & 0.45$^{+0.03}_{-0.02}$                & 0.46$^{+0.02}_{-0.02}$\\
10, 10 & 5.0$^{+0.4}_{-0.4}$$\times10^{-2}$    & 0.16$^{+0.01}_{-0.01}$                & 0.17$^{+0.01}_{-0.01}$                & 0.18$^{+0.01}_{-0.01}$                & 0.19$^{+0.01}_{-0.01}$                & 9.5$^{+0.5}_{-0.5}$$\times10^{-2}$    & 0.21$^{+0.01}_{-0.01}$                & 0.21$^{+0.01}_{-0.01}$\\
10, 5  & 1.2$^{+0.3}_{-0.2}$$\times10^{-2}$    & 6.8$^{+0.6}_{-0.5}$$\times10^{-2}$    & 7.0$^{+0.6}_{-0.6}$$\times10^{-2}$    & 8.1$^{+0.6}_{-0.6}$$\times10^{-2}$    & 7.8$^{+0.6}_{-0.6}$$\times10^{-2}$    & 4.4$^{+0.5}_{-0.4}$$\times10^{-2}$    & 8.9$^{+0.6}_{-0.6}$$\times10^{-2}$    & 9.2$^{+0.6}_{-0.6}$$\times10^{-2}$\\
5, 5   & 3.6$^{+0.6}_{-0.6}$$\times10^{-3}$    & 3.7$^{+0.2}_{-0.2}$$\times10^{-2}$    & 3.0$^{+0.2}_{-0.2}$$\times10^{-2}$    & 4.4$^{+0.2}_{-0.2}$$\times10^{-2}$    & 3.4$^{+0.2}_{-0.2}$$\times10^{-2}$    & 2.1$^{+0.1}_{-0.1}$$\times10^{-2}$    & 4.6$^{+0.2}_{-0.2}$$\times10^{-2}$    & 4.7$^{+0.2}_{-0.2}$$\times10^{-2}$\\
\end{tabular}
\end{ruledtabular}
\caption[Sensitive hypervolume measurements for individual search pipelines and combined pipeline configurations]{\label{tab:sens_pastro}Search pipeline sensitive hypervolumes for different mass combinations, using a probability of astrophysical origin greater than or equal to 0.5 and a false alarm rate less than 2 per day as the detection threshold. The combined columns refer to the combined sensitive hypervolume, requiring that at least one search pipeline detects a given event. The asterisk ($^*$) refers to the combination of the cWB, GstLAL, MBTA, PyCBC-broad, and PyCBC-BBH searches, and the dagger ($^\dagger$) refers to the combined sensitivity after including our search in this list.}
\end{table*}}

{\renewcommand{\arraystretch}{1.5}%
\begin{table*}[]
\begin{ruledtabular}
\begin{tabular}{c cccccc}
 & cWB & GstLAL & MBTA & PyCBC-broad & PyCBC-BBH & This Work\\
\hline
Total &       18,118 &     29,303 &        27,060 &      26,589 &             31,392 &           23,238\\ 
\hline
Unique$^*$ &  1,000 &        1,155 &         1,796 &       270 &                2,460 &            -\\ 
Unique$^\dagger$ &   684 &        946 &           1,370 &       236 &                2,043 &            2,612 
\\
\end{tabular}
\end{ruledtabular}
\caption[Total detected event counts and counts of uniquely detected events for each search pipeline]{\label{tab:num_detections}Total counts of detected injections in O3 for each search pipeline (top row) and the number of uniquely detected events for each search pipeline before ($^*$) and after ($^\dagger$), including the results of our search pipeline (bottom two rows). The detection threshold used is a probability of astrophysical origin greater than or equal to 0.5 and a false alarm rate less than 2 per day. These results demonstrate the significant and unique detection contributions of our work, as well as our support for detecting events previously identified by a single search pipeline.}
\end{table*}}

The calculation of the sensitive hypervolume is described in Refs.~\cite{gwtc3, gwtc3_inj_set}, where it quantifies the sensitivity of a search to a population of sources uniformly distributed in comoving volume and source-frame time. The expected number of detections, $\hat{N}$, for a search is~\cite{gwtc3}
\begin{equation}\label{eq:vt}
\hat{N} = \braket{VT} R,
\end{equation}
where $R$ is the rate of signals per unit volume per unit observing time. The thresholds we use to compute sensitivity results are $p_{\rm astro}\geq0.5$ and an FAR of less than 2 per day. For our search, we assign to an injection the most significant trigger within $\pm1$ seconds of the merger time. We compute the sensitive hypervolume for the mass combinations used in the GWTC-3 paper~\cite{gwtc3}, together with three additional mass pairs with a primary mass of 50$\,$M$_{\odot}$, to parametrize the sensitivity across the parameter space and to observe trends in sensitive regions for each search.

\begin{figure*}[p]
\includegraphics[width=0.832\linewidth]{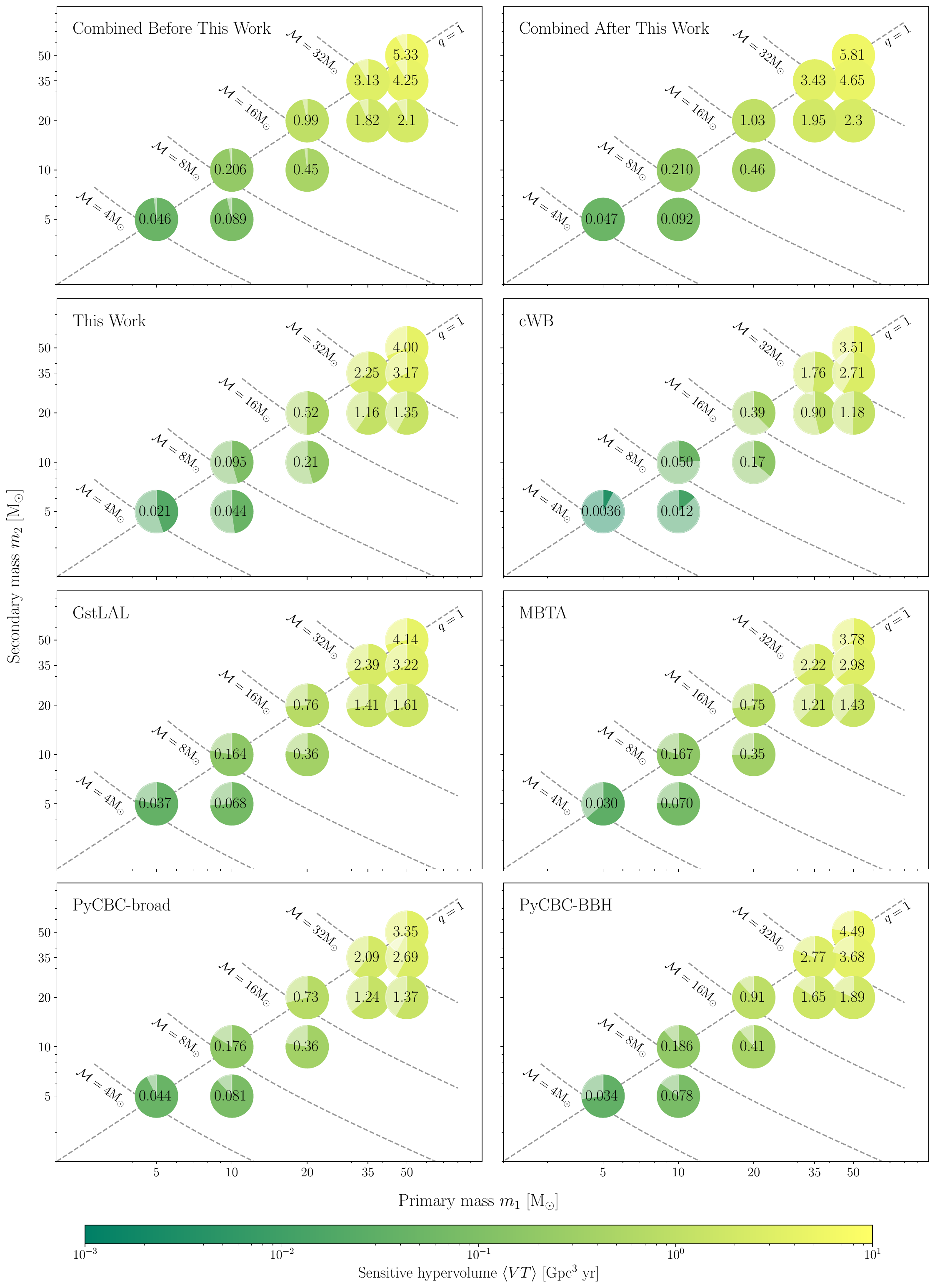}
\caption[Visual representation of individual and combined pipeline sensitive hypervolumes]{\label{fig:sens_pastro}Sensitive hypervolume, $\braket{VT}$, from the O3 observing run for our search pipeline and the searches presented in GWTC-3~\cite{gwtc3}. Additionally, we include two $\braket{VT}$ estimates for injections identified by at least one search pipeline, both before and after considering our own search pipeline, to demonstrate the significant, unique contributions of our analysis. The detection threshold is $p_{\rm astro}\geq0.5$ and FAR $<$ 2 per day, and we compute the sensitive hypervolume at component mass pairs that act as the central point of a log-normal distribution with a standard deviation of 0.1$\,$M$_{\odot}$. Each $\braket{VT}$ estimate is marked with a pie chart where the darker region indicates the fraction of the combined $\braket{VT}$ that is recovered. The color of the darker region of each point is defined by the value of the $\braket{VT}$, as given by the color scale. The $\braket{VT}$ values and mass pairs are the same as those given in Table~\ref{tab:sens_pastro}.}
\end{figure*}

In Table~\ref{tab:sens_pastro}, we report the sensitive hypervolume at each mass combination used to parametrize the search space. Figure~\ref{fig:sens_pastro} shows the variation in the sensitive hypervolume between each search across the BBH parameter space. To calculate the sensitive hypervolume around each mass pair, injections are weighted so that they follow a log-normal distribution around the central mass with a standard deviation of 0.1$\,$M$_{\odot}$, and we assume that the component spins are isotropically distributed with uniformly distributed magnitudes~\cite{gwtc3}. In addition to computing the sensitivity of each search included in the GWTC-3 paper, we calculate combined sensitivity metrics with and without our search pipeline to assess the impact of our search on combined detection efforts.

Relative to PyCBC-BBH, our search is less sensitive overall. Among the all-source pipelines (GstLAL, MBTA, PyCBC-broad), our search pipeline is comparable at higher masses but degrades for signals at or below a source-frame chirp mass of $\sim$25$\,$M$_{\odot}$ and an expected network SNR of $\sim$15. This low-mass sensitivity gap does not imply an intrinsic limitation of the method, as we have shown in past work that we can detect BNS signals~\cite{McLeod:bns}. Instead, it likely reflects that the training dataset and configuration choices did not optimize sensitivity over the vast BBH space.

For mass pairs with a 50$\,$M$_{\odot}$ primary component, our sensitivity appears to drop as the mass ratio, $q$, decreases relative to other searches; however, the size of our injection set prevents us from independently investigating mass-ratio effects and overall low-mass sensitivity.

Including our search increases the combined sensitive hypervolume, notably from the 20-20$\,$M$_{\odot}$ bin and above (Table~\ref{tab:sens_pastro}). As expected, our contribution to the combined sensitive volume decreases at lower masses, where our sensitivity is lower.

Table~\ref{tab:num_detections} provides the total number of detections from each search pipeline during the O3 observing run, as well as the count of unique detections for each pipeline in the combined detection scenarios, both with and without our search pipeline included. GWTC-3~\cite{gwtc3} requires that injections detected by cWB are also detected by another search, but we do not impose that criterion here. Our search yields the most unique detections among the overall population of injected signals. The following section, Section~\ref{sec:pycbc_comparison}, demonstrates that this significant result is a feature of our deep-learning ranking statistic approach.

\subsection{\label{sec:pycbc_comparison}Comparison with a BBH-only PyCBC search}

To determine whether the significant, unique contributions of our search are due to the template bank or to our ranking statistic, we ran the PyCBC search pipeline using our template bank over weeks 4, 6, and 8 and compared the detected signal populations. Here, we are using the PyCBC-broad version of the pipeline, which does not include weighting higher-chirp-mass triggers, unlike the PyCBC-BBH pipeline.

Because estimates of $p_{\rm astro}$ were not available in the PyCBC configuration we used, we set a detection threshold of FAR $<$ 2 per month for both pipelines, consistent with the open public alert threshold used for real-time detections in O3~\cite{gwtc3}. The change from using $p_{\rm astro}$ as a detection threshold to using FAR is not a concern, as we find that the injection recovery rates are consistent with the counts observed in Table~\ref{tab:num_detections}.

There are 10,472 total injections in the set analyzed by PyCBC using our bank, and the detection counts for each search using FAR as the threshold are shown in Table~\ref{tab:num_detections_pycbc}. We observe that the PyCBC ranking statistic identifies more injections than our search, as expected given our low total sensitivity. Despite this, the number of uniquely detected events in the combined detections configuration for our search pipeline (348) is significantly higher than PyCBC using our bank (23), indicating that our deep-learning ranking statistic is the reason we can detect a unique population of signals, rather than the choice of template bank. Based on this finding, the large number of uniquely identified events in our search, outlined in Table~\ref{tab:num_detections}, is due to our unique search method rather than our choice of template bank.

{\renewcommand{\arraystretch}{1.5}%
\begin{table*}[!t]
\begin{ruledtabular}
\begin{tabular}{c ccccccc}
 & cWB & GstLAL & MBTA & PyCBC-broad & PyCBC-BBH & This Work & PyCBC-Our Bank\\
\hline
Total &              2,427 &      3,835 &    3,271 &       3,122 &         3,751 &       2,916 &       3,369\\
\hline
Unique$^*$ &         75 &         188 &      183 &         28 &            249 &         -   &        -\\
Unique$^\dag$ &   66 &         164 &      129 &         26 &            207 &         348 &        -\\
Unique$^\ddag$ &  67 &         179 &      180 &         15 &            211 &         -   &        23\\
\end{tabular}
\end{ruledtabular}
\caption[Total detected event counts and counts of uniquely detected events for each search pipeline when comparing our search to the PyCBC search using our template bank]{\label{tab:num_detections_pycbc}Total count of detected injections for each search pipeline in weeks 4, 6 and 8 of the third observing run (top row). Number of uniquely detected events per search pipeline in three different combinations of searches: the five GWTC-3 search pipelines ($^*$), the five GWTC-3 search pipelines plus this work ($^\dag$), and the five GWTC-3 search pipelines plus PyCBC-Our Bank ($^\ddag$). The PyCBC-Our Bank column refers to our PyCBC analysis using the template bank used by our search pipeline. The detection threshold used here is a false alarm rate of less than 1 per 2 months. The larger quantity of unique detections from our work compared to PyCBC-Our Bank demonstrates that our work's unique detection ability stems from our deep-learning ranking statistic rather than the template bank.}
\end{table*}}

\begin{figure}[t]
\includegraphics[width=\linewidth]{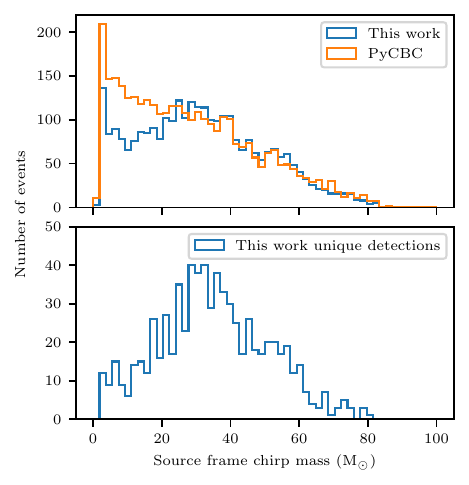}
\caption[Comparison of events detected over different source frame chirp masses for our search and the PyCBC search using our template bank]{\label{fig:mine_vs_pycbc}Comparison of detected injection counts over source-frame chirp mass between PyCBC (run using our template bank) and our search pipeline, for injections in weeks 4, 6, and 8 of the O3 observing run (top). Distribution of events detected by our search pipeline and not detected by PyCBC with our template bank (bottom).}
\end{figure}

Figure~\ref{fig:mine_vs_pycbc} demonstrates that the difference in total detection counts between our search pipeline and PyCBC with our template bank is due to our lack of sensitivity at low masses. Conversely, when directly comparing the populations of detected events using only our search pipeline and PyCBC with our template bank, our search identifies 726 unique events. The distribution of source-frame chirp masses for our search's 726 uniquely detected events is also plotted in Fig.~\ref{fig:mine_vs_pycbc}. This distribution closely matches that of all detected events, suggesting that an increase in sensitivity at lower masses will likely yield even more uniquely detected events; however, this is not guaranteed. Large unique detection counts further suggest that, despite detecting a consistent number of events at higher chirp masses, the search pipelines are not detecting the same events, which may become very important when conducting searches for real gravitational wave signals.

\subsection{\label{sec:sens_over_time}Sensitivity over time}

\begin{figure*}[!t]
\includegraphics[width=\linewidth]{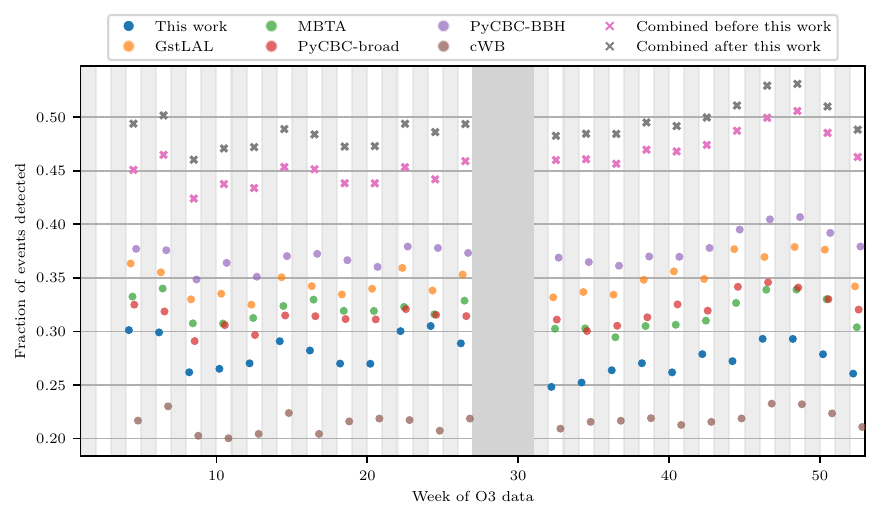}
\caption[Search sensitivity over time for the O3 observing run for individual and combined search pipeline configurations]{\label{fig:sens_over_time}Search sensitivities in terms of the fraction of detected events over the course of the O3 observing run. The detection threshold for each search pipeline is a probability of astrophysical origin greater than or equal to 0.5 and a false alarm rate less than 2 per day. The combined sensitivity metrics (crosses) refer to the scenario where at least one search pipeline detects an event, both before and after our search pipeline is included. These combined sensitivity metrics demonstrate the impact of the unique detections made by our search pipeline and that our sensitivity is consistent over time.}
\end{figure*}

\begin{figure}[h]
\includegraphics[width=\linewidth]{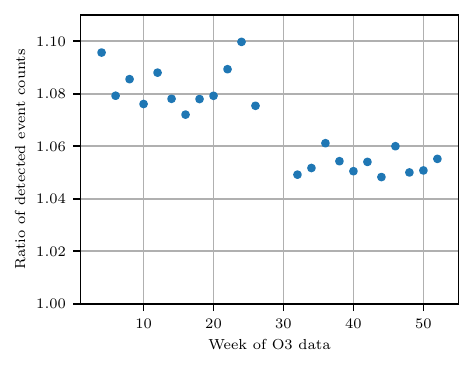}
\caption[Improvement in sensitivity over O3 for the combined search pipelines before and after including our search pipeline]{\label{fig:any_ratio-sens_over_time}Ratio of the fraction of detected events from Figure~\ref{fig:sens_over_time} for the combined detection scenario with and without including our search pipeline. Values greater than one indicate that including our search pipeline increases the number of events detected by at least one search pipeline. The decrease after week 30 suggests that our search pipeline makes fewer unique detections in O3b than in O3a.}
\end{figure}

Since we need to train a deep learning model for our search, it is essential to determine how often the model will need to be retrained due to drift in the detector sensitivity. It is also necessary to consider the change in search sensitivity due to the instrument upgrades performed between O3a and O3b~\cite{aLIGO:2020wna}.

To visualize how our model's sensitivity changes over time, we compute the fraction of detected signals for each injection run week we analyze. Figure~\ref{fig:sens_over_time} presents the fraction of signals detected with $p_{\rm astro}\geq0.5$ and FAR $<$ 2 per day for each pipeline and the combined detection scenarios before and after including our search pipeline. We use the fraction of recovered events as the metric, as in Ref.~\cite{aframe}, to separate the detector sensitivity from the search-pipeline sensitivities. We observe no significant or consistent change in sensitivity over time and note that our results follow the general trends in sensitivity observed during the O3 observing run.

Going from O3a to O3b, there appears to be a drop in sensitivity for our search relative to other searches, as well as a drop in the increase between the combined detection configurations. We look at this further in Figure~\ref{fig:any_ratio-sens_over_time}, where we observe a significant decrease between O3a and O3b in the ratio of total detected events before and after including our search in the combined detection configurations. However, the minimum increase in the number of detected events by including our search is only $4.74\,\%$, which remains a significant addition. The large spread in Fig.~\ref{fig:any_ratio-sens_over_time} makes it challenging to verify the sensitivity to unique events over time; however, training our deep-learning model on real noise from more extended periods throughout O3 is expected to stabilise the sensitivity.

From these results, it is clear that retraining would not be necessary on a weekly or monthly time scale with our current training setup. We could use noise from a more extended period to train the model, rather than only the first week of O3, to reduce variation in the model's sensitivity over time. Other deep-learning gravitational wave searches have demonstrated satisfactory performance with longer training time scales~\cite{aresgw,nagarajan2025}. Our decrease in unique detection counts between O3a and O3b suggests that retraining between observing runs, when detector upgrades are performed, is important for improving our search performance over time.

\section{\label{sec:search}Offline search}

In this section, we conduct an offline search for BBH gravitational wave signals in O3 data. We present the results of this search in terms of candidate events identified by our search pipeline, as well as those candidates that we do not identify, which have been identified by previous existing offline searches. These previous offline searches include the GWTC~\cite{gwtc2,gwtc21,gwtc3}, OGC~\cite{ogc3,ogc4}, IAS~\cite{ias_o3a,ias_o3b,ias_hom}, PyCBC-KDE~\cite{pycbc_kde}, cWB~\cite{cwb_offline}, AresGW~\cite{aresgw} and Aframe~\cite{aframe_o3_search} catalogs. The OGC, IAS, and PyCBC-KDE catalogs are produced using matched-filter searches; cWB is an unmodeled burst-search pipeline; and AresGW and Aframe are deep-learning searches that identify candidates from detector strain data.

With a candidate identification threshold of $p_{\rm astro}\geq0.5$ and FAR less than 2 per day, we identified 33 candidate events over O3. This includes 23 in O3a and 10 in O3b. Of these, 31 are consistent with the most up-to-date GWTC catalogs; one candidate (GW190605\_025957) was consistent only with the IAS higher-order modes search, IAS-HOM~\cite{ias_hom}, and one was a newly identified candidate that has not been previously reported (GW190929\_091722). For candidates identified by our search, we report the $p_{\rm astro}$, FAR, and network SNR values, as well as the corresponding values from the original offline search that led to their identification, in Table~\ref{tab:found-cands}. In the case of GWTC being the original search, we include the results from the pipeline (cWB, MBTA, GstLAL, PyCBC or PyCBC-BBH) with the highest $p_{\rm astro}$, or highest SNR if more than one search pipeline shares the highest $p_{\rm astro}$. The candidates we jointly detect with existing offline searches are presented in more detail in Section~\ref{sec:old-cands}, and we explore the newly detected candidate in Section~\ref{sec:new-cands}. We investigate the candidates that our search pipeline failed to identify in Section~\ref{sec:missed-cands}.

\subsection{\label{sec:old-cands}Existing candidate detections}

We see in Table~\ref{tab:found-cands} that the candidate events identified by our search strongly overlap with the results of the GWTC catalogs, including one of five NSBH candidates, GW190814\_211039, that we discuss further in Section~\ref{sec:found-gw190814}. In addition, we provide supporting evidence for the identification of one candidate previously identified only by the IAS-HOM search~\cite{ias_hom} and one new candidate unique to our search. We discuss these two candidates further in Sections~\ref{sec:found-ias_hom} and~\ref{sec:new-cands}, respectively.

We find two candidate events, GW190731\_140936 and GW200216\_220804, that are detected with $p_{\rm astro}$ values significantly higher than in their original searches. Both of these candidates were later reported with higher $p_{\rm astro}$ values in the OGC-4 catalog~\cite{ogc4}, but these updated values are still lower than those assigned by our search. We also find that our $p_{\rm astro}$ for GW200216\_220804 is higher despite having a worse FAR than reported in GWTC-3. To understand these behaviors, further investigation is required using injection campaigns to compare $p_{\rm astro}$ values across search pipelines.

\subsubsection{\label{sec:found-gw190814}GW190814\_211039}

Our searched mass space includes a subset of the theoretical neutron star population as our lower mass limit is 2$\,$M$_{\odot}$. GW190814\_211039 is one of five potential NSBH candidates in O3 and is the only NSBH candidate we identify in the GWTC catalogs. Our results indicate that we identify it with lower $p_{\rm astro}$ and higher FAR than the search pipelines in the GWTC catalogs, as expected given our observed low sensitivity to low-mass events. It is the lowest-chirp-mass source to pass our detection threshold, and its significance is likely due to its high network SNR of 22.6.

Improvements to NSBH sensitivity would require training samples that represent a population of NSBH signals rather than just low mass BBH signals, as well as a $p_{\rm astro}$ classifier that considers independent populations of BNS, NSBH and BBH candidates.

{\renewcommand{\arraystretch}{1.25}%
\begin{table*}[]
\begin{tabular}{m{3cm} m{1.75cm} | >{\centering}m{1cm} >{\centering}m{2cm} >{\centering}m{1.25cm} | >{\centering}m{1cm} >{\centering}m{2cm} m{1.25cm}<{\centering}}
\hline \hline
& & \multicolumn{3}{c|}{This Work} & \multicolumn{3}{c}{Original Search}\\
Name & Catalog & $p_{\rm astro}$ & FAR (yr$^{-1}$) & SNR & $p_{\text{astro}}$ & FAR (yr$^{-1}$) & SNR\\
\hline
GW190408\_181802 & GWTC-2.1 & 1.00  & 1.5$\times$10$^{-5}$ & 14.0 & 1.00 & $<$1.0$\times$10$^{-5}$ & 14.7\\
GW190412\_053044 & GWTC-2.1 & 1.00  & 7.8$\times$10$^{-6}$ & 17.2 & 1.00 & $<$1.0$\times$10$^{-5}$ & 19.0\\
GW190413\_134308 & GWTC-2.1 & 0.84  & 2.0                  & 9.7  & 0.99 & 0.34                    & 10.3\\
GW190421\_213856 & GWTC-2.1 & 0.99  & 0.067                & 10.1 & 1.00 & 0.0028                  & 10.5\\
GW190503\_185404 & GWTC-2.1 & 1.00  & 5.9$\times$10$^{-4}$ & 12.5 & 1.00 & 0.013                   & 12.8\\
GW190517\_055101 & GWTC-2.1 & 1.00  & 1.1$\times$10$^{-3}$ & 10.2 & 1.00 & 0.11                    & 11.3\\
GW190519\_153544 & GWTC-2.1 & 1.00  & 0.0036               & 13.8 & 1.00 & 7.0$\times$10$^{-5}$    & 13.7\\
GW190521\_074359 & GWTC-2.1 & 1.00  & 2.6$\times$10$^{-9}$ & 22.9 & 1.00 & $<$1.0$\times$10$^{-5}$ & 24.4\\
GW190527\_092055 & GWTC-2.1 & 0.82  & 4.3                  & 8.2  & 0.85 & 0.23                    & 8.7 \\
GW190602\_175927 & GWTC-2.1 & 1.00  & 0.0020               & 12.0 & 1.00 & 3.0$\times$10$^{-4}$    & 12.6\\
GW190605\_025957 & IAS-HOM  & 0.78  & 4.6                  & 8.7  & 0.88 & 1.7                     & 9.5 \\
GW190706\_222641 & GWTC-2.1 & 1.00  & 1.6$\times$10$^{-5}$ & 12.4 & 1.00 & 0.34                    & 12.6\\
GW190707\_093326 & GWTC-2.1 & 1.00  & 0.0090               & 12.3 & 1.00 & $<$1.0$\times$10$^{-5}$ & 13.2\\
GW190727\_060333 & GWTC-2.1 & 1.00  & 5.8$\times$10$^{-5}$ & 11.6 & 1.00 & $<$1.0$\times$10$^{-5}$ & 12.1\\
GW190728\_064510 & GWTC-2.1 & 1.00  & 0.0041               & 12.4 & 1.00 & $<$1.0$\times$10$^{-5}$ & 13.4\\
GW190731\_140936 & GWTC-2.1 & 0.97  & 0.52                 & 8.4  & 0.83 & 1.9                     & 7.8 \\
GW190803\_022701 & GWTC-2.1 & 0.99  & 0.21                 & 8.7  & 0.97 & 0.39                    & 8.7 \\
GW190814\_211039$^\dagger$ & GWTC-2.1 & 0.68  & 0.17                 & 22.6 & 1.00 & $<$1.0$\times$10$^{-5}$ & 22.2\\
GW190828\_063405 & GWTC-2.1 & 1.00  & 3.1$\times$10$^{-8}$ & 15.1 & 1.00 & $<$1.0$\times$10$^{-5}$ & 16.3\\
GW190828\_065509 & GWTC-2.1 & 0.88  & 1.3                  & 10.0 & 1.00 & 3.5$\times$10$^{-5}$    & 11.1\\
GW190915\_235702 & GWTC-2.1 & 1.00  & 3.7$\times$10$^{-6}$ & 12.4 & 1.00 & $<$7.0$\times$10$^{-5}$ & 13.1\\
GW190929\_012149 & GWTC-2.1 & 0.88  & 1.8                  & 9.2  & 0.87 & 0.16                    & 10.1\\
GW190929\_091722 & This Work& 0.63  & 27                   & 6.7  &\dots & \dots                   &\dots\\

GW191109\_010717 & GWTC-3 & 1.00  & 1.3$\times$10$^{-4}$ & 16.1 & $>$0.99& 0.001                   & 15.8\\
GW191129\_134029 & GWTC-3 & 1.00  & 7.3$\times$10$^{-4}$ & 12.5 & $>$0.99& $<$1.0$\times$10$^{-5}$ & 13.3\\
GW191215\_223052 & GWTC-3 & 0.89  & 0.56                 & 10.1 & $>$0.99& $<$1.0$\times$10$^{-5}$ & 10.9\\
GW191222\_033537 & GWTC-3 & 1.00  & 2.3$\times$10$^{-4}$ & 11.8 & $>$0.99& $<$1.0$\times$10$^{-5}$ & 12.0\\
GW200129\_065458 & GWTC-3 & 1.00  & 3.2$\times$10$^{-8}$ & 24.3 & $>$0.99& $<$1.0$\times$10$^{-5}$ & 26.5\\
GW200208\_130117 & GWTC-3 & 0.92  & 0.53                 & 9.5  & $>$0.99& 3.1$\times$10$^{-4}$    & 10.8\\
GW200209\_085452 & GWTC-3 & 0.99  & 0.044                & 9.4  & 0.97   & 12                      & 9.7 \\
GW200216\_220804 & GWTC-3 & 0.85  & 1.5                  & 8.8  & 0.77   & 0.35                    & 9.4 \\
GW200224\_222234 & GWTC-3 & 1.00  & 7.0$\times$10$^{-8}$ & 17.9 & $>$0.99& $<$8.2$\times$10$^{-5}$ & 19.2\\
GW200311\_115853 & GWTC-3 & 1.00  & 1.3$\times$10$^{-6}$ & 15.7 & $>$0.99& $<$1.0$\times$10$^{-5}$ & 17.7\\
\hline \hline
\end{tabular}
\caption[Candidates from an offline search over O3 with $p_{\rm astro}\geq0.5$]{\label{tab:found-cands}Search pipeline trigger properties of events identified by our search during the third observing run with a probability of astrophysical origin ($p_{\rm astro}$) greater than or equal to 0.5 and an FAR less than 2 per day. Original search refers to the first offline search to identify an event with $p_{\rm astro}\geq0.5$, as reported in the catalog column; in the case of GWTC, it is the results of the most significant individual search pipeline for that event. GW190814\_211039 is indicated with a $^\dagger$ due to it being a potential NSBH candidate, although it cannot be ruled out as a BBH candidate.}
\end{table*}}

\subsubsection{\label{sec:found-ias_hom}GW190605\_025957}

This candidate was first reported in the IAS higher-order modes search~\cite{ias_hom} and is jointly identified for the first time here. This is a reasonably significant result, as offline searches independent of the GWTC catalogs have previously identified 49 additional candidates during O3. Of these, only 10 have been jointly identified by independent searches~\cite{beyond_gwtc3}, where we do not consider targeted searches (e.g., IAS-HOM) as independent of other searches using the same software framework (IAS-O3a and IAS-O3b). The parameter estimation results for this candidate suggest that it was produced by a high-mass BBH system~\cite{ias_hom,beyond_gwtc3}. However, the candidate has a low SNR, resulting in unconstrained priors.

To test whether this candidate is caused by a glitch or data quality issue, we examine transient noise events and data quality issues, and perform a signal consistency test around the time of the event. To identify excess-power transient events, we use the Omicron software package~\cite{omicron}, which reports one transient in the Livingston detector. This transient has an SNR of 6.05 and a frequency of 41.8$\,$Hz, and occurs within 30$\,$ms of the trigger produced by our search, indicating that this candidate may not be astrophysical. We use the results from the \textsc{iDQ} framework~\cite{iDQ,idq_o3a_zenodo} - a machine-learning framework that correlates auxiliary channels with the detector strain - to assess data quality, and we found no significant trigger values near the candidate's event time. Lastly, we employ a reduced $\chi^2$ signal consistency test from the PyCBC search pipeline~\cite{Nitz2018} to determine if the distribution of signal power over frequency bins matches the expectation from the template. We compute the chi-square statistic over a number of bins that depends on the template parameters of the template that produced the trigger in our search pipeline. The formula can be found in the OGC-4 data release\footnote{\url{https://github.com/gwastro/4-ogc}}~\cite{ogc4}. To compute the PyCBC reweighted SNR statistic, the signal SNR would be down-weighted by a factor of 0.908 in the Hanford detector and 0.987 in the Livingston detector by considering Eqn. 5 in Ref.~\cite{Nitz2018}. These checks suggest that further investigation of the nature of this candidate may be required to determine if it originates from an astrophysical signal.

\subsection{\label{sec:new-cands}New candidate event}

\begin{figure}[t]
\includegraphics[width=\linewidth]{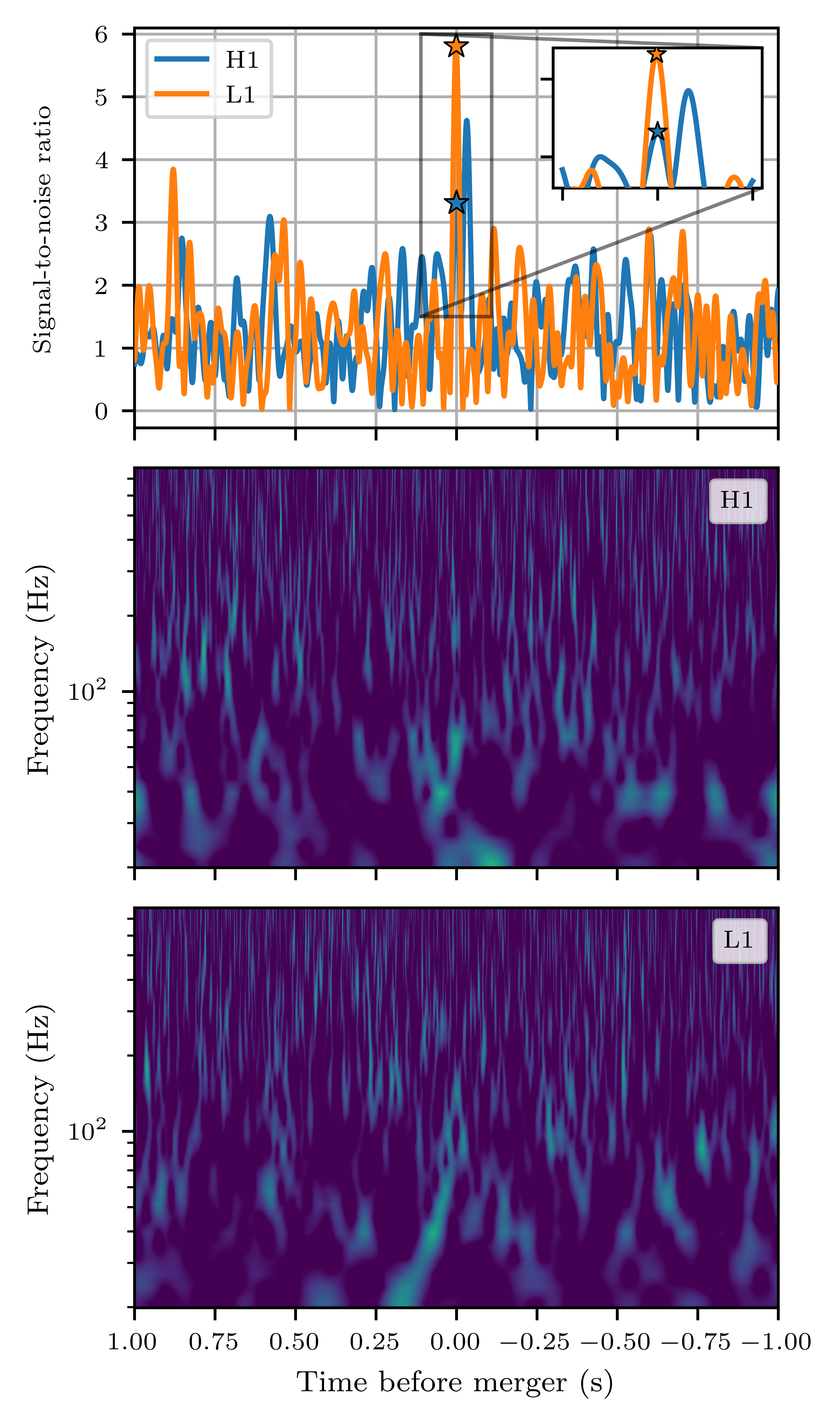}
\caption[SNR time series and spectrograms for Hanford and Livingston detector data surrounding our unique candidate GW190929\_091722]{\label{fig:gw190929_data}SNR time series and time-frequency spectrograms of the strain data for our new candidate event GW190929\_091722. The stars on the SNR time series represent the values from the trigger that identified the event, highlighting a larger SNR peak in the Hanford detector (H1) outside the light-travel time from the peak in the Livingston detector (L1).}
\end{figure}

We also report, for the first time, the identification of a candidate event, GW190929\_091722, at GPS time 1253783860.8. We present the Hanford and Livingston SNR time series and time-frequency spectrograms around the event time in Figure~\ref{fig:gw190929_data}. Existing searches have not detected this candidate event, and there are no coincident candidates in the sub-threshold candidate releases as part of the GWTC~\cite{gwtc21_candidates,gwtc3_candidates} and OGC~\cite{ogc3,ogc4} catalogs.

In Figure~\ref{fig:gw190929_data}, we include a star symbol to highlight the SNR peaks, identified by our peak finding algorithm, associated with the trigger that identifies this candidate. We see that the SNR peak in the Hanford detector is below 4. In contrast, existing multi-detector matched-filter searches consider only trigger times when more than one detector peaks above an SNR of 4. Since we do not train our model on low SNR and single detector samples, and we tune our $p_{\rm astro}$ model on a set of injections with a minimum network SNR of 6, there is a reasonable amount of uncertainty in our $p_{\rm astro}$ calculation. Additionally, the Hanford peak identified by our algorithm is not associated with the higher SNR feature that occurs approximately 30$\,$ms post-merger. A common concern in developing deep learning models is explainability, as it is often unclear which aspects of the input data the model relies on to make its predictions. In this case, the candidate’s SNR time series underscores the need to examine how deep learning models behave in practice when detecting coincident signals. In particular, we must ensure that the model relies only on astrophysically relevant features.

As we did for GW190605\_025957, we conducted a series of data quality checks to validate the data around the candidate event. A search for transient noise events with Omicron returned no triggers with an SNR greater than 5 at a frequency that would appear in the SNR time series. Two transients were identified by Omicron in the Hanford detector at $\sim$446$\,$Hz and $\sim$1345$\,$Hz with SNRs of 5.03 and 5.56, respectively, and were labeled as occurring at least 30$\,$ms after the candidate's merger time. Since these transients occur at high frequencies, have low SNRs, and appear after the candidate merger time, we do not expect them to interfere with identifying this candidate, which was done using a high-mass, short-duration template. We also find that no significant trigger values were produced by the \textsc{iDQ} framework around this event~\cite{idq_o3a_zenodo}. Lastly, the reduced $\chi^2$ was $<$ 1 in both the Hanford and Livingston detectors, which would result in no scaling of the SNR to get the PyCBC reweighted SNR statistic~\cite{Nitz2018}. These findings suggest that there are no data-quality or transient-noise issues with the data surrounding this candidate event.

\subsubsection{\label{sec:pe}Parameter estimation}

For our new candidate event, we based our parameter-estimation analysis on Ref.~\cite{beyond_gwtc3} and used the \textsc{Asimov} software framework~\cite{asimov,asimov_version} to manage and execute the parameter-estimation workflows. We used \textsc{BayesWave}~\cite{2015CQGra..32m5012C,PhysRevD.91.084034} to produce on-source PSD estimates in both the Hanford and Livingston detectors. These PSDs were used with \textsc{Bilby}~\cite{bilby_paper} for model selection with the dynesty nested sampler~\cite{dynesty}, together with the IMRPhenomXPHM waveform model~\cite{Pratten:2020ceb}. We also used the \textsc{PESummary} package~\cite{pesummary} for plotting and analysis of the posterior estimates produced by \textsc{Bilby}. We provide the necessary \textsc{Asimov} configuration files, as well as the resulting posterior samples, as part of an associated data release on GitHub\footnote{\url{https://github.com/damonbeveridge/new_BBH_O3_candidate-UWA}}.

\begin{figure*}[!t]
\includegraphics[width=\linewidth]{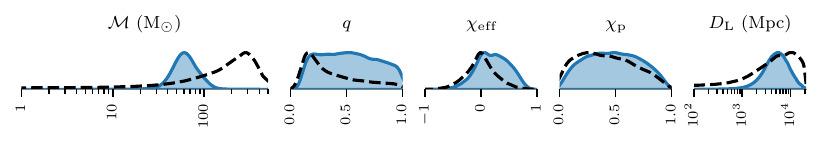}
\caption[A subset of parameter estimation posteriors for our unique candidate event GW190929\_091722]{\label{fig:pe_with_priors}Parameter estimation results for our new candidate event, GW190929\_091722. From left to right are the source frame chirp mass, $\mathcal{M}$; the mass ratio, $q$; the effective dimensionless spin parameter, $\chi_{\rm eff}$; the precessing dimensionless spin parameter, $\chi_{\rm p}$; and the luminosity distance, $D_{\rm L}$. The blue region represents the posterior probability distributions from our analysis, and the black dotted line represents the prior distributions.}
\end{figure*}

\begin{figure}[]
\includegraphics[width=\linewidth]{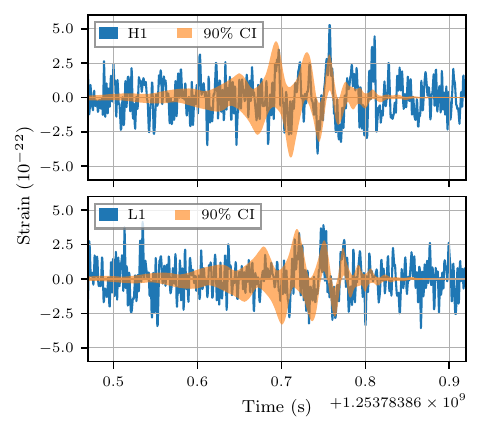}
\caption[Strain data, with the 90\% credible region waveform from parameter estimation, for our new candidate GW190929\_091722]{\label{fig:waveform_reconstruction}Detector strain data and Bilby posterior waveform 90\% credible region for the gravitational-wave candidate event GW190929\_091722 observed by the LIGO Hanford (H1, top panel) and LIGO Livingston (L1, bottom panel) detectors. Times are shown relative to September 29, 2019 at 09:17:22 UTC. The strain time series (blue) is filtered with a 35–350 Hz bandpass filter to suppress large fluctuations outside the detectors’ most sensitive frequency band. The orange area shows the 90\% credible interval for the waveform reconstruction based on posterior samples from a Bilby parameter-estimation study.}
\end{figure}

We present the results of this analysis for a subset of the source parameters in Table~\ref{tab:event_params} and Figure~\ref{fig:pe_with_priors}. We see that this candidate is consistent with a high mass BBH binary with a total mass, $M$, of $171^{+172}_{-58}\,$M$_{\odot}$, which is a larger median than any GWTC-3~\cite{gwtc3} candidate and the third highest median compared with the other searches in the Beyond GWTC-3 parameter estimation catalog~\cite{beyond_gwtc3}. Figure~\ref{fig:waveform_reconstruction} compares the detector strain data around the time of the event with the 90\% credible regions of the Bilby reconstructed waveforms for each of the Hanford and Livingston detectors.

Existing black hole formation models present the case that black holes cannot be expected to be formed through stellar collapse if the black hole mass is in the range of $65<m<120$~\cite{Mehta:2021fgz,Fowler:1964zz,Barkat:1967zz,Fryer:2000my,Belczynski:2016jno,Spera:2017fyx,Stevenson:2019rcw,Costa:2020xbc,Farmer:2020xne,Renzo:2020lwl}, and that some other formation channel is required. In this mass range, we expect a progenitor to encounter pair-instability supernovae, leaving no remnant behind. Our posterior results for GW190929\_091722 suggest that the primary component has a 51.6\% probability of having a mass within this range, and the secondary component has a 34.7\% probability. Additionally, our primary component has a 44.5\% probability of having a mass greater than 120$\,$M$_{\odot}$.

Our parameter estimation results indicate that, if astrophysical, this candidate could lie outside the mass parameter space on which our deep learning model is trained, thereby affecting the significance with which we recover it. To account for this, an intermediate-mass black hole search should be conducted to verify our search results for this candidate event. The LVK Collaboration performed an offline IMBH search in O3~\cite{lvk_imbh_o3}, and found no candidates coincident with our new candidate. We see no other notable results from this parameter estimation analysis, mainly due to the low SNR and short duration of the candidate.

{\renewcommand{\arraystretch}{1.5}%
\begin{table*}[]
\begin{ruledtabular}
\begin{tabular}{ccccccccccc}
$M$\,(M$_{\odot}$) & $\mathcal{M}$\,(M$_{\odot}$) & $m_1$\,(M$_{\odot}$) & $m_2$\,(M$_{\odot}$) & $\chi_{\mathrm{eff}}$ & $D_{\mathrm{L}}$\,(Gpc) & $z$ & $M_{\mathrm{f}}$\,(M$_{\odot}$) & $\chi_{\mathrm{f}}$ & $\Delta\Omega$(deg$^2$) & SNR\\
\hline
171$^{+172}_{-58}$ & 68.0$^{+53}_{-25}$ & 114$^{+162}_{-47}$ & 57.5$^{+48}_{-33}$ & 0.21$^{+0.45}_{-0.47}$ & 7.02$^{+8.25}_{-4.20}$ & 1.03$^{+0.90}_{-0.54}$ & 164$^{+169}_{-55}$ & 0.76$^{+0.16}_{-0.25}$ & 15000 & 5.55$^{+0.78}_{-1.37}$\\
\end{tabular}
\end{ruledtabular}
\caption[Parameter estimates for our unique candidate event, GW190929\_091722]{\label{tab:event_params}Median and 90\% symmetric credible intervals for select source parameters of our new candidate event, GW190929\_091722. The columns show the total mass of the binary, $M$; the chirp mass, $\mathcal{M}$; the component masses, $m_1$ and $m_2$; the effective dimensionless spin parameter, $\chi_{\rm eff}$; the luminosity distance, $D_{\rm L}$; the redshift, $z$; the final mass, $M_f$; the dimensionless final spin $\chi_{\rm f}$; the sky localization area, $\Delta\Omega$; and the network matched-filter SNR. All quoted results for masses are in the source frame.}
\end{table*}}

\subsection{\label{sec:missed-cands}Missed candidates}

Equally important as the candidates we detect are those we fail to recover. Examining missed events highlights the current limitations of our pipeline. We display the candidate events that our search fails to identify throughout O3a and O3b in Table~\ref{tab:missed_o3a} and Table~\ref{tab:missed_o3b}, respectively. We include the results of our search pipeline in these tables when a trigger is produced with an FAR of less than 2 per day.

In Section~\ref{sec:hopeless-cands} below, we discuss the candidate events that are not included in Tables~\ref{tab:missed_o3a} and~\ref{tab:missed_o3b} due to design choices of our search pipeline that prevent us from being able to detect these events. Section~\ref{sec:remaining-cands} discusses the remaining candidate events that are considered detectable by our search pipeline, and we give reasoning for why these candidates are not detected.

{\renewcommand{\arraystretch}{1.25}%
\begin{table*}[]
\begin{tabular}{m{3cm} m{1.75cm} | >{\centering}m{1.5cm} >{\centering}m{1.5cm} | >{\centering}m{1cm} >{\centering}m{2cm} >{\centering}m{1.25cm} | >{\centering}m{1cm} >{\centering}m{2cm} m{1.25cm}<{\centering}}
\hline \hline
& & & & \multicolumn{3}{c|}{This Work} & \multicolumn{3}{c}{Original Search}\\
Name & Catalog & $\mathcal{M}$\,(M$_{\odot}$) & q & $p_{\rm astro}$ & FAR (yr$^{-1}$) & SNR & $p_{\rm astro}$ & FAR (yr$^{-1}$) & SNR\\
\hline
GW190403\_051519 & GWTC-2.1           & $34.0^{+15.1}_{-8.4}$   & $0.23^{+0.57}_{-0.12}$ & \dots & \dots & \dots & 0.61  & 7.7                     & 8.0  \\
GW190404\_142514 & OGC-4              & $13.8^{+2.2}_{-1.9}$    & $0.56^{+0.36}_{-0.27}$ & \dots & \dots & \dots & 0.50  & 50                      & 7.8  \\
GW190413\_052954 & GWTC-2.1           & $24.6^{+5.5}_{-4.1}$    & $0.73^{+0.24}_{-0.31}$ & 0.27  & 44    & 8.7   & 0.93  & 0.82                    & 8.5  \\
GW190426\_082124 & AresGW             & $28.9^{+8.7}_{-7.4}$    & $0.70^{+0.26}_{-0.34}$ & \dots & \dots & \dots & 0.50  & 20                      & -    \\
GW190426\_190642 & GWTC-2.1           & $76.0^{+19.1}_{-17.4}$  & $0.76^{+0.22}_{-0.48}$ & \dots & \dots & \dots & 0.75  & 4.1                     & 9.6  \\
GW190427\_180650 & OGC-4              & $7.3^{+0.6}_{-0.6}$     & $0.54^{+0.38}_{-0.26}$ & \dots & \dots & \dots & 0.53  & 50                      & 8.9  \\
GW190511\_125545 & AresGW             & $27.2^{+9.7}_{-7.6}$    & $0.61^{+0.34}_{-0.33}$ & \dots & \dots & \dots & 1.00  & 0.27                    & -    \\
GW190511\_163209 & IAS-HOM            & $39.2^{+35.5}_{-23.6}$  & $0.27^{+0.59}_{-0.16}$ & \dots & \dots & \dots & 0.61  & 12                      & 9.5  \\
GW190512\_180714 & GWTC-2.1           & $14.6^{+1.3}_{-1.0}$    & $0.54^{+0.36}_{-0.18}$ & 0.24  & 18    & 11.2  & 1.00  & $<$1.1$\times$10$^{-4}$ & 12.4 \\
GW190521\_030229 & GWTC-2.1           & $69.2^{+17.0}_{-10.6}$  & $0.59^{+0.33}_{-0.38}$ & 0.06  & 36    & 14.4  & 1.00  & 0.0013                  & 13.6 \\
GW190523\_085933 & AresGW             & $23.2^{+18.0}_{-12.0}$  & $0.38^{+0.53}_{-0.21}$ & \dots & \dots & \dots & 0.68  & 20                      & -    \\
GW190524\_134109 & IAS-HOM            & $46.0^{+16.8}_{-13.8}$  & $0.66^{+0.30}_{-0.44}$ & 0.05  & 444   & 8.0   & 0.85  & 2.3                     & 8.2  \\
GW190530\_030659 & IAS-HOM            & $21.4^{+3.7}_{-3.7}$    & $0.61^{+0.33}_{-0.35}$ & 0.04  & 578   & 7.8   & 0.63  & 11                      & 8.4  \\
GW190530\_133833 & IAS-HOM            & $58.6^{+23.6}_{-20.6}$  & $0.59^{+0.37}_{-0.39}$ & \dots & \dots & \dots & 0.53  & 17                      & 8.4  \\
GW190604\_103812 & IAS-HOM            & $55.4^{+31.5}_{-25.5}$  & $0.35^{+0.48}_{-0.21}$ & \dots & \dots & \dots & 0.68  & 7.7                     & 8.2  \\
GW190607\_083827 & AresGW             & $31.6^{+7.4}_{-6.6}$    & $0.77^{+0.21}_{-0.31}$ & \dots & \dots & \dots & 0.99  & 6.5                     & -    \\
GW190614\_134749 & AresGW             & $26.3^{+9.6}_{-7.6}$    & $0.60^{+0.35}_{-0.33}$ & \dots & \dots & \dots & 0.99  & 4.6                     & -    \\
GW190615\_030234 & IAS-HOM            & $49.8^{+12.5}_{-10.9}$  & $0.75^{+0.22}_{-0.37}$ & \dots & \dots & \dots & 0.75  & 5.0                     & 8.6  \\
GW190704\_104834$^\ddagger$ & IAS-O3a & $4.2^{+0.1}_{-0.1}$     & $0.52^{+0.38}_{-0.29}$ & \dots & \dots & \dots & 0.81  & 0.36                    & 8.9  \\
GW190705\_164632 & AresGW             & $33.6^{+11.2}_{-10.3}$  & $0.57^{+0.38}_{-0.32}$ & \dots & \dots & \dots & 0.51  & 49                      & -    \\
GW190707\_083226 & IAS-O3a            & $35.9^{+8.9}_{-7.9}$    & $0.63^{+0.32}_{-0.34}$ & \dots & \dots & \dots & 0.94  & 0.043                   & 8.3  \\
GW190711\_030756 & IAS-O3a            & $31.1^{+6.7}_{-6.4}$    & $0.34^{+0.36}_{-0.21}$ & \dots & \dots & \dots & 0.93  & 0.089                   & 9.0  \\
GW190718\_160159 & IAS-O3a            & $7.0^{+0.7}_{-0.7}$     & $0.68^{+0.28}_{-0.41}$ & \dots & \dots & \dots & 0.53  & 2.1                     & 8.4  \\
GW190719\_215514 & GWTC-2.1           & $23.5^{+6.5}_{-4.0}$    & $0.55^{+0.40}_{-0.41}$ & 0.10  & 177   & 8.3   & 0.92  & 0.63                    & 8.0  \\
GW190720\_000836 & GWTC-2.1           & $8.9^{+0.5}_{-0.8}$     & $0.53^{+0.36}_{-0.24}$ & 0.12  & 65    & 10.1  & 1.00  & 0.094                   & 11.6 \\
GW190805\_105432 & OGC-4              & $8.8^{+0.8}_{-0.8}$     & $0.52^{+0.37}_{-0.20}$ & \dots & \dots & \dots & 0.51  & 50                      & 8.1  \\
GW190805\_211137 & GWTC-2.1           & $31.9^{+8.8}_{-6.3}$    & $0.68^{+0.28}_{-0.33}$ & \dots & \dots & \dots & 0.95  & 0.63                    & 8.3  \\
GW190806\_033721 & IAS-HOM            & $42.4^{+15.2}_{-12.8}$  & $0.49^{+0.44}_{-0.30}$ & \dots & \dots & \dots & 0.86  & 2.2                     & 8.3  \\
GW190814\_192009 & IAS-O3a            & $50.3^{+17.2}_{-13.9}$  & $0.71^{+0.26}_{-0.40}$ & \dots & \dots & \dots & 0.64  & 1.5                     & 8.0  \\
GW190821\_124821$^\ddagger$ & IAS-O3a & $4.8^{+0.3}_{-0.2}$     & $0.28^{+0.28}_{-0.11}$ & \dots & \dots & \dots & 0.60  & 1.4                     & 8.8  \\
GW190904\_104631 & AresGW             & $26.1^{+15.8}_{-15.3}$  & $0.61^{+0.34}_{-0.34}$ & \dots & \dots & \dots & 0.72  & 14                      & -    \\
GW190906\_054335 & IAS-O3a            & $25.7^{+8.4}_{-8.0}$    & $0.63^{+0.33}_{-0.36}$ & \dots & \dots & \dots & 0.61  & 1.4                     & 7.9  \\
GW190910\_012619 & IAS-O3a            & $9.6^{+0.8}_{-0.9}$     & $0.12^{+0.04}_{-0.02}$ & \dots & \dots & \dots & 0.58  & 1.5                     & 8.2  \\
GW190916\_200658 & GWTC-2.1           & $26.9^{+8.2}_{-5.4}$    & $0.55^{+0.40}_{-0.32}$ & 0.44  & 30    & 7.9   & 0.66  & 6900                    & 8.2  \\
GW190917\_114630$^\dagger$ & GWTC-2.1 & $3.7^{+0.2}_{-0.2}$     & $0.21^{+0.32}_{-0.09}$ & \dots & \dots & \dots & 0.77  & 0.66                    & 9.5  \\
GW190920\_113516$^\ddagger$ & IAS-O3a & $3.8^{+0.2}_{-0.2}$     & $0.52^{+0.38}_{-0.29}$ & \dots & \dots & \dots & 0.57  & 1.8                     & 8.6  \\
GW190924\_021846 & GWTC-2.1           & $5.8^{+0.2}_{-0.2}$     & $0.58^{+0.32}_{-0.30}$ & 0.07  & 61    & 11.9  & 1.00  & $<$1.0$\times$10$^{-5}$ & 13.0 \\
GW190926\_050336 & GWTC-2.1           & $24.4^{+9.0}_{-4.9}$    & $0.50^{+0.42}_{-0.26}$ & \dots & \dots & \dots & 0.54  & 1.1                     & 9.0  \\
GW190930\_133541 & GWTC-2.1           & $8.5^{+0.5}_{-0.5}$     & $0.49^{+0.43}_{-0.27}$ & 0.45  & 11    & 10.1  & 1.00  & 0.012                   & 10.0\\
\hline \hline
\end{tabular}
\caption[Search results for O3a candidates reported by other searches that our search identifies with $p_{\rm astro}\leq0.5$]{\label{tab:missed_o3a}Properties of missed events during the O3a observing run that are detectable according to Section~\ref{sec:hopeless-cands}. Original search refers to the first offline search that identifies an event. For GWTC-3 candidates, the results are from the most significant (highest $p_{\rm astro}$) search pipeline for that event. The dagger ($^\dagger$) indicates a candidate with masses consistent with an NSBH system, and a double dagger ($^\ddagger$) indicates a candidate that is possibly consistent with originating from an NSBH or BBH source. Ellipses (\dots) indicate that our search pipeline does not produce a trigger with an FAR $<$ 2 per day.}
\end{table*}}

{\renewcommand{\arraystretch}{1.25}%
\begin{table*}[]
\begin{tabular}{m{3cm} m{1.75cm} | >{\centering}m{1.5cm} >{\centering}m{1.5cm} | >{\centering}m{1cm} >{\centering}m{2cm} >{\centering}m{1.25cm} | >{\centering}m{1cm} >{\centering}m{2cm} m{1.25cm}<{\centering}}
\hline \hline
& & & & \multicolumn{3}{c|}{This Work} & \multicolumn{3}{c}{Original Search}\\
Name & Catalog & $\mathcal{M}$\,(M$_{\odot}$) & $q$ & $p_{\rm astro}$ & FAR (yr$^{-1}$) & SNR & $p_{\rm astro}$ & FAR (yr$^{-1}$) & SNR\\
\hline
GW191103\_012549 & GWTC-3             & $8.3^{+0.7}_{-0.6}$     & $0.67^{+0.29}_{-0.37}$ & \dots & \dots & \dots & 0.94  & 0.46                    & 9.3  \\
GW191105\_143521 & GWTC-3             & $7.8^{+0.6}_{-0.5}$     & $0.72^{+0.24}_{-0.31}$ & \dots & \dots & \dots &$>$0.99& 0.14                    & 10.7 \\
GW191113\_071753 & GWTC-3             & $10.7^{+1.1}_{-1.0}$    & $0.20^{+0.49}_{-0.09}$ & \dots & \dots & \dots & 0.68  & 26                      & 9.2  \\
GW191113\_103541 & IAS-HOM            & $36.3^{+13.3}_{-9.7}$   & $0.20^{+0.18}_{-0.07}$ & \dots & \dots & \dots & 0.76  & 4.8                     & 8.7  \\
GW191117\_023843 & IAS-O3b            & $38.3^{+9.7}_{-7.9}$    & $0.26^{+0.17}_{-0.14}$ & \dots & \dots & \dots & 0.56  & 8.3                     & 8.0  \\
GW191126\_115259 & GWTC-3             & $8.7^{+1.0}_{-0.7}$     & $0.69^{+0.28}_{-0.35}$ & \dots & \dots & \dots & 0.70  & 3.2                     & 8.5  \\
GW191127\_050227 & GWTC-3             & $29.9^{+11.7}_{-9.1}$   & $0.47^{+0.47}_{-0.35}$ & \dots & \dots & \dots & 0.74  & 4.1                     & 8.7  \\
GW191204\_110529 & GWTC-3             & $19.8^{+3.6}_{-3.2}$    & $0.73^{+0.24}_{-0.35}$ & \dots & \dots & \dots & 0.74  & 3.3                     & 8.9  \\
GW191224\_043228 & OGC-4              & $9.3^{+0.9}_{-0.8}$     & $0.51^{+0.39}_{-0.22}$ & \dots & \dots & \dots & 0.87  & 7.7                     & 8.4  \\
GW191228\_085854 & IAS-O3b            & $6.3^{+0.2}_{-0.2}$     & $0.57^{+0.20}_{-0.27}$ & \dots & \dots & \dots & 0.67  & 4.0                     & 9.0  \\
GW191228\_195619 & IAS-HOM            & $96.0^{+40.4}_{-31.8}$  & $0.46^{+0.32}_{-0.22}$ & \dots & \dots & \dots & 0.67  & 8.3                     & 10.5 \\
GW191230\_180458 & GWTC-3             & $36.5^{+8.2}_{-5.6}$    & $0.77^{+0.20}_{-0.34}$ & 0.48  & 5.4   & 10.0  & 0.96  & 0.42                    & 9.9  \\
GW200106\_134123 & OGC-4              & $29.7^{+7.4}_{-7.2}$    & $0.63^{+0.32}_{-0.30}$ & 0.45  & 20    & 7.5   & 0.69  & 17                      & 7.4  \\
GW200109\_195634 & IAS-O3b            & $44.9^{+12.7}_{-11.7}$  & $0.72^{+0.25}_{-0.37}$ & 0.32  & 26    & 8.0   & 0.81  & 1.1                     & 8.0  \\
GW200115\_042309$^\dagger$ & GWTC-3   & $2.4^{+0.1}_{-0.1}$     & $0.24^{+0.43}_{-0.10}$ & 0.02  & 227   & 10.4  &$>$0.99& $<$1.0$\times$10$^{-5}$ & 11.5 \\
GW200128\_022011 & GWTC-3             & $32.0^{+7.5}_{-5.5}$    & $0.80^{+0.18}_{-0.30}$ & \dots & \dots & \dots &$>$0.99& 0.0043                  & 9.9  \\
GW200129\_114245 & OGC-4              & $43.1^{+16.0}_{-15.6}$  & $0.45^{+0.44}_{-0.24}$ & \dots & \dots & \dots & 0.53  & 25                      & 7.9  \\
GW200202\_154313 & GWTC-3             & $7.5^{+0.2}_{-0.2}$     & $0.72^{+0.24}_{-0.31}$ & \dots & \dots & \dots &$>$0.99& $<$1.0$\times$10$^{-5}$ & 11.3 \\
GW200208\_211609 & AresGW             & $21.0^{+35.6}_{-14.5}$  & $0.72^{+0.24}_{-0.31}$ & \dots & \dots & \dots & 0.55  & 18                      & -    \\
GW200208\_222617 & GWTC-3             & $19.8^{+10.5}_{-5.2}$   & $0.21^{+0.67}_{-0.16}$ & \dots & \dots & \dots & 0.70  & 4.8                     & 7.9  \\
GW200210\_005122 & OGC-4              & $6.4^{+0.5}_{-0.4}$     & $0.67^{+0.29}_{-0.33}$ & \dots & \dots & \dots & 0.74  & 25                      & 8.3  \\
GW200210\_092254$^\ddagger$ & GWTC-3  & $6.6^{+0.4}_{-0.4}$     & $0.12^{+0.05}_{-0.04}$ & \dots & \dots & \dots & 0.54  & 7.7                     & 8.9  \\
GW200210\_100022 & IAS-O3b            & $29.1^{+8.9}_{-7.3}$    & $0.17^{+0.20}_{-0.06}$ & \dots & \dots & \dots & 0.52  & 10                      & 7.8  \\
GW200219\_094415 & GWTC-3             & $27.6^{+5.6}_{-3.8}$    & $0.77^{+0.21}_{-0.32}$ & 0.07  & 67    & 10.1  &$>$0.99& 9.9$\times$10$^{-4}$    & 10.7 \\
GW200220\_061928 & GWTC-3             & $62.2^{+22.8}_{-15.1}$  & $0.73^{+0.24}_{-0.41}$ & \dots & \dots & \dots & 0.62  & 6.8                     & 7.5  \\
GW200220\_124850 & GWTC-3             & $28.2^{+7.3}_{-5.1}$    & $0.74^{+0.23}_{-0.33}$ & \dots & \dots & \dots & 0.83  & 1800                    & 8.2  \\
GW200225\_060421 & GWTC-3             & $14.2^{+1.5}_{-1.4}$    & $0.73^{+0.23}_{-0.28}$ & \dots & \dots & \dots &$>$0.99& $<$8.8$\times$10$^{-4}$ & 13.1 \\
GW200225\_075134 & IAS-O3b            & $42.8^{+9.0}_{-7.9}$    & $0.77^{+0.20}_{-0.31}$ & \dots & \dots & \dots & 0.60  & 6.7                     & 8.3  \\
GW200301\_211019 & IAS-HOM            & $14.5^{+2.1}_{-2.1}$    & $0.63^{+0.31}_{-0.40}$ & \dots & \dots & \dots & 0.56  & 14                      & 8.6  \\
GW200304\_172806 & IAS-HOM            & $52.6^{+21.9}_{-19.6}$  & $0.49^{+0.45}_{-0.35}$ & \dots & \dots & \dots & 0.66  & 9.1                     & 8.3  \\
GW200305\_084739 & OGC-4              & $24.0^{+5.6}_{-4.7}$    & $0.75^{+0.22}_{-0.35}$ & \dots & \dots & \dots & 0.59  & 50                      & 7.6  \\
GW200306\_093714 & GWTC-3             & $17.5^{+3.5}_{-3.0}$    & $0.53^{+0.40}_{-0.33}$ & \dots & \dots & \dots & 0.81  & 410                     & 8.5  \\
GW200308\_173609 & GWTC-3             & $34.2^{+43.8}_{-17.5}$  & $0.39^{+0.48}_{-0.29}$ & \dots & \dots & \dots & 0.86  & 2.4                     & 8.0  \\
GW200316\_235947$^\ddagger$ & IAS-O3b & $4.2^{+0.2}_{-0.2}$     & $0.57^{+0.33}_{-0.37}$ & \dots & \dots & \dots & 0.52  & 11                      & 8.3  \\
GW200318\_191337 & OGC-4              & $31.9^{+8.5}_{-6.5}$    & $0.70^{+0.26}_{-0.34}$ & \dots & \dots & \dots & 0.97  & 2                       & 7.8  \\
GW200322\_091133 & GWTC-3             & $15.0^{+29.5}_{-4.0}$   & $0.29^{+0.63}_{-0.26}$ & \dots & \dots & \dots & 0.62  & 450                     & 9.0\\
\hline \hline
\end{tabular}
\caption[Search results for O3b candidates reported by other searches that our search identifies with $p_{\rm astro}\leq0.5$]{\label{tab:missed_o3b}Properties of missed events during the O3b observing run that are detectable according to Section~\ref{sec:hopeless-cands}. Original search refers to the first offline search that identifies an event. For GWTC-3 candidates, the results are from the most significant (highest $p_{\rm astro}$) search pipeline for that event. The dagger ($^\dagger$) indicates a candidate with masses consistent with an NSBH system, and a double dagger ($^\ddagger$) indicates a candidate that is possibly consistent with originating from an NSBH or BBH source. Ellipses (\dots) indicate that our search pipeline does not produce a trigger with an FAR $<$ 2 per day.}
\end{table*}}

\subsubsection{\label{sec:hopeless-cands}Hopeless candidates}

Excluded from Tables~\ref{tab:missed_o3a} and~\ref{tab:missed_o3b} are 11 events due to no coincident Hanford and Livingston detector data, including six in O3a and three in O3b from the GWTC-3 catalog, as well as one in each of the observation runs from the OGC catalog. Additionally, we miss GW190911\_195101 because it occurs in a data segment with a duration less than 1024 seconds, which we exclude from our search as per Section~\ref{sec:data}.

\begin{figure}[t]
\includegraphics[width=\linewidth]{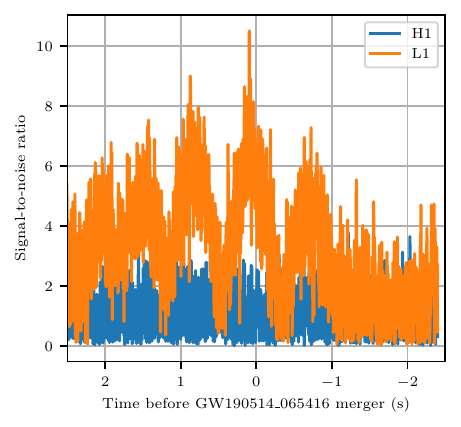}
\caption[SNR time series produced by a glitch that dominated a GWTC-3 candidate event in our search pipeline]{\label{fig:glitch_dominated}SNR time series for the template that produced the trigger in our search pipeline nearest to the event GW190514\_065416. This template has a detector frame chirp mass of 2.9$\,$M$_{\odot}$. In contrast, GW190514\_065416 has an estimated detector-frame chirp mass of 48.6$\,$M$_{\odot}$, demonstrating that our search pipeline is susceptible to long-duration templates broadening the temporal response of a short-duration glitch, resulting in missed detections.}
\end{figure}

Since our search pipeline clusters SNR triggers across the entire template bank using our peak-finding algorithm, we find that it has missed four events due to glitch interference. We miss these events when long-duration templates produce high SNR responses to glitches over long timescales that dominate our peak-finding algorithm, even after the glitch has ceased to be present in the strain. The candidates affected by this include GW190514\_065416, GW190818\_232544, GW191204\_171526, and GW191219\_163120, the latter of which has source properties consistent with an NSBH event. An example of this behavior is shown in Figure~\ref{fig:glitch_dominated}, which displays the SNR time series of the highest-SNR template identified by our peak-finding algorithm for GW191204\_171526. The PyCBC search pipeline overcomes this by performing peak-finding and producing a ranking statistic across 15 separate clusters of their template bank, then marginalizing over these clusters by selecting the trigger with the most significant ranking statistic~\cite{ogc1}. This feature can be integrated into our search pipeline in the future, particularly as we expand it to identify BNS, NSBH, and BBH candidates. From a preliminary analysis comparing the recovered SNR to the injected SNR of our injection campaign during the O3 observing run, this feature may make over a thousand injections now detectable by our search, potentially providing a meaningful increase in our sensitivity.

Since our search involves a matched-filter analysis, we lose the ability to identify events in the first 200 seconds and the last 24 seconds of a data segment, as outlined in Section~\ref{sec:data}. The 200-second cuts account for four missed events from the GWTC catalogs - GW190701\_203306, GW190725\_174728, GW200214\_223307, GW200316\_215756 - and the 24-second cuts account for one event - GW190513\_205428. Suppose we were to develop our search pipeline further to identify short-duration templates within the first 200 seconds of a data segment. It may then be possible to detect the four missed events, as they should be short enough signals within the detector sensitivity bands based on their inferred masses~\cite{gwtc3}. The 24-second window at the end of a data segment was chosen as a conservative value, and should be able to be reduced in the future. Reducing this window could help our search identify GW190513\_205428, which exists approximately 21 seconds before the end of a data segment.

\subsubsection{\label{sec:remaining-cands}Detectable candidates}

In this section, we relate trends from our injection sensitivity tests to the population of remaining missed events in our O3 search. We discuss only the events listed in Tables~\ref{tab:missed_o3a} and~\ref{tab:missed_o3b} for which our search pipeline has the opportunity to detect candidates. This includes 39 events in O3a and 36 events in O3b.

The most notable lack of sensitivity we observed in our injection studies (Section~\ref{sec:search-sens}) is at lower masses. More specifically, we observe a drop-off in sensitivity relative to the other matched-filter searches below a source-frame chirp mass of 25$\,$M$_{\odot}$ and a network SNR of 15. By considering the median source frame chirp masses from parameter estimation~\cite{gwtc2,gwtc21,beyond_gwtc3,ogc4,ias_o3a,ias_o3b} and the original search SNRs of the events in Tables~\ref{tab:missed_o3a} and~\ref{tab:missed_o3b}, this accounts for 18 of 39 events in O3a and 19 of 36 in O3b, including four (three) potential NSBH candidates in O3a (O3b). Extending this selection to events where the parameter estimation results indicate that the 90\% credible region for the source-frame chirp mass is less than 25$\,$M$_{\odot}$ results in 28 of 39 missed events in O3a and 26 of 36 in O3b.

Of the remaining 21 missed candidates throughout O3, nine candidates have a mass ratio, $q$, less than 0.5. In Section~\ref{sec:search-sens}, we observed the possibility that our search has a reduced sensitivity for higher mass events with a low mass ratio. However, these sensitivity results are not conclusive, so we cannot conclusively attribute this to the failure to detect these events. Of these low mass ratio events, GW190604\_103812, GW191113\_103541 and GW191228\_195619 have a median primary component mass greater than 100$\,$M$_{\odot}$, which is outside our trained parameter space. The remaining three events with $q<0.5$ in each O3a and O3b have primary component masses with 90\% credible intervals extending above 100$\,$M$_{\odot}$ as well.

There remain eight (six) missed events in O3a (O3b) for which we have not provided reasoning through a link to our injection sensitivity results. We find that in O3a, GW190426\_190642 has a median primary-component mass above our training parameter space. Additionally, GW190530\_133833 and GW200220\_061928 have posterior support within the 90\% credible interval for a primary component mass exceeding the limits of our training parameter space.

Of the remaining 11 events throughout O3 that we do not detect, our search pipeline does produce triggers with an FAR of less than 2 per day for three candidates. Our search identified GW191230\_180458 with a $p_{\rm astro}$ of 0.48, and GW200109\_195634 with a $p_{\rm astro}$ of 0.32, so we do not consider missing these events problematic. We produced a trigger for GW190524\_134109 with a $p_{\rm astro}$ of 0.05; however, this event has been identified only by the IAS-HOM~\cite{ias_hom}.

Three of the remaining eight candidates that we do not produce triggers for are supported by detections from more than one independent search pipeline: GW190607\_083827~\cite{aresgw,cwb_offline}, GW200128\_022011~\cite{gwtc3}, and GW200318\_191337~\cite{ogc4,cwb_offline}. Since the remaining five candidates cannot be explicitly treated as astrophysical due to low SNR and a lack of supporting detections across independent searches, we are not concerned about not detecting them.

While we have demonstrated a link between our injection sensitivity results and the population of missed events, it does not necessarily explain why there are so many missed events. It is essential to note that, although one or more searches have identified a candidate, it does not necessarily mean that the candidate event is guaranteed to originate from an astrophysical source. If we only consider candidate events from the LVK's GWTC-3 catalog~\cite{gwtc3}, we have identified 31 of 79 candidates throughout O3, with only 63 of those being detectable based on constraints outlined in Section~\ref{sec:hopeless-cands}. Additionally, our injection studies revealed a large number of unique detections, resulting in less overlap between the detected populations of our search pipeline and the others, which may be translated into our search for real events.

\section{\label{sec:conc}Conclusion}

In this work, we conducted an injection sensitivity study and an offline search for BBH gravitational waves during O3. Unlike traditional pipelines that rely on analytical metrics to identify signals, we built a hybrid search pipeline that uses a deep learning classifier to predict from the SNR time series produced by matched filtering. The current version of the search pipeline demonstrates comparable sensitivity to existing searches for high-stellar-mass black holes, yielding a large number of unique detections that significantly enhance the combined search sensitivity of existing pipelines. Our observed low sensitivity at low masses is not a concern for future applications of this search method, given past research demonstrating its capabilities at low masses; therefore, a revised training and optimization approach is required. We demonstrate that this search method can detect 31 previously identified gravitational wave candidate events from the GWTC catalogs, as well as one candidate from the IAS higher-order modes search and a new candidate that has not been previously reported. Several more candidates have been identified with high significance by the GWTC, OGC, IAS, and AresGW offline searches, and we have provided a link between our injection sensitivity results and the population of missed candidates.

We have discussed several areas for improvement in our search pipeline as we progress to developing a search that can identify BNS, NSBH and BBH candidates together. These include optimizing training datasets and training metrics to ensure consistent sensitivity levels across the parameter space, improving sensitivity over long observing runs so that models do not need to be retrained regularly, and updating our $p_{\rm astro}$ model to model our ranking statistic and source populations more accurately and to classify between source types. For the training datasets, we note potential areas for improvement, including the selection of templates for generating SNR time-series samples and the parameter distributions of injection waveforms.

There are also more specific pipeline improvements that could be adapted from existing search pipelines, such as clustering regions of the template bank to identify BBH events in the presence of loud glitches, which resulted in four missed candidates. Additional potential changes include adjusting our SNR time-series peak-finding algorithm to account for calibration uncertainties in the light travel time between detectors and to require coincident triggers between detectors. Lastly, extending our search to single-detector and other two-detector configurations will increase our capacity to identify events beyond times when the Hanford and Livingston detectors are operating in coincidence. Furthermore, adding a three-detector search with the Virgo detector could increase confidence in detections.

\begin{acknowledgments}

We thank Siddharth Soni for their assistance in providing Omicron glitch identification results for the O3 observing run. This work was performed on the OzSTAR national facility at Swinburne University of Technology. The OzSTAR program receives funding in part from the Astronomy National Collaborative Research Infrastructure Strategy (NCRIS) allocation provided by the Australian Government, and from the Victorian Higher Education State Investment Fund (VHESIF) provided by the Victorian Government. This material is based upon work supported by NSF's LIGO Laboratory, which is a major facility fully funded by the National Science Foundation. The authors are grateful for computational resources provided by the LIGO Laboratory and supported by National Science Foundation Grants PHY-0757058 and PHY-0823459.
\end{acknowledgments}

\bibliography{apssamp}

\end{document}